\documentclass[twocolumn,trackchanges]{aastex7}

\usepackage{CJK}
\usepackage[utf8]{inputenc}
\usepackage{graphicx}
\usepackage{tabularx}
\usepackage{amsmath}
\usepackage{soul}

\DeclareMathAlphabet{\mathsc}{OT1}{cmr}{m}{sc}
\def\testbx{bx}%
\DeclareRobustCommand{\ion}[2]{%
\relax\ifmmode
\ifx\testbx\f@series
{\mathbf{#1\,\mathsc{#2}}}\else
{\it{#1\,\mathsc{#2}}}\fi
\else\textup{#1\,{\mdseries\textsc{#2}}}%
\fi}
\received{Month XX, 2026}
\revised{Month XX, 2026}
\accepted{Month XX, 2026}
\shorttitle{SN~2025dys observational properties}
\shortauthors{Jiewei~Zhao et al.}

\begin{document}
\title{SN~2025dys: A luminous supernova explosion from a partially stripped helium star}
\author[0009-0009-8633-8582]{Jiewei~Zhao}
\affiliation{South-Western Institute for Astronomy Research, Yunnan University, Kunming, Yunnan 650500, People's Republic of China}
\affiliation{Yunnan Key Laboratory of Survey Science, Yunnan University, Kunming, Yunnan 650500, People's Republic of China}
\email{meow.jiewei.zhao@gmail.com}
\author[0000-0001-7225-2475]{Brajesh~Kumar}
\affiliation{South-Western Institute for Astronomy Research, Yunnan University, Kunming, Yunnan 650500, People's Republic of China}
\affiliation{Yunnan Key Laboratory of Survey Science, Yunnan University, Kunming, Yunnan 650500, People's Republic of China}
\email{brajesh@ynu.edu.cn, brajesharies@gmail.com}
\author[0009-0009-4872-1134]{Gitika~Rameshan}
\affiliation{Indian Institute of Astrophysics, II Block, Koramangala, Bengaluru-560034, Karnataka, India}
\affiliation{Academy of Scientific and Innovative Research (AcSIR), Ghaziabad, Uttar Pradesh, 201002, India}
\email{gitika.rameshan@iiap.res.in}
\author[0000-0002-6688-0800]{Devendra~K.~Sahu}
\affiliation{Indian Institute of Astrophysics, II Block, Koramangala, Bengaluru-560034, Karnataka, India}
\email{dks@iiap.res.in}
\author[0000-0003-4111-5958]{Bin-Bin~Zhang}
\affiliation{School of Astronomy and Space Science, Nanjing University, Nanjing 210093, People's Republic of China}
\email{bbzhang@nju.edu.cn}
\author[0009-0005-1517-6089]{Kamil~Nadaf}
\affiliation{School of Astronomy and Space Science, Nanjing University, Nanjing 210093, People's Republic of China}
\email{kamilnadaf.phy@gmail.com}
\author[0000-0002-8109-7152]{Guowang~Du}
\affiliation{South-Western Institute for Astronomy Research, Yunnan University, Kunming, Yunnan 650500, People's Republic of China}
\affiliation{Yunnan Key Laboratory of Survey Science, Yunnan University, Kunming, Yunnan 650500, People's Republic of China}
\email{dugking@ynu.edu.cn}
\author[0009-0001-1479-9090]{Wenyu~Li}
\affiliation{South-Western Institute for Astronomy Research, Yunnan University, Kunming, Yunnan 650500, People's Republic of China}
\affiliation{Yunnan Key Laboratory of Survey Science, Yunnan University, Kunming, Yunnan 650500, People's Republic of China}
\email{yanyustarfallrain@gmail.com}
\author[0009-0000-7773-553X]{Zehui~Peng}
\affiliation{School of Physics and Astronomy, Beijing Normal University, Beijing 100875, China}
\email{zehui-peng@foxmail.com}
\author[0000-0002-2510-6931]{Jinghua~Zhang}
\affiliation{South-Western Institute for Astronomy Research, Yunnan University, Kunming, Yunnan 650500, People's Republic of China}
\affiliation{Yunnan Key Laboratory of Survey Science, Yunnan University, Kunming, Yunnan 650500, People's Republic of China}
\email{zhang_jh@ynu.edu.cn}
\author[0000-0003-4121-5684]{Shiyan~Zhong}
\affiliation{South-Western Institute for Astronomy Research, Yunnan University, Kunming, Yunnan 650500, People's Republic of China}
\affiliation{Yunnan Key Laboratory of Survey Science, Yunnan University, Kunming, Yunnan 650500, People's Republic of China}
\email{zhongsy@ynu.edu.cn}
\author[0000-0003-0394-1298]{Xiangkun Liu}
\affiliation{South-Western Institute for Astronomy Research, Yunnan University, Kunming, Yunnan 650500, People's Republic of China}
\affiliation{Yunnan Key Laboratory of Survey Science, Yunnan University, Kunming, Yunnan 650500, People's Republic of China}
\email{liuxk@ynu.edu.cn}
\author[0000-0003-2091-622X]{Avinash~Singh}
\affiliation{The Oskar Klein Centre, Department of Astronomy, Stockholm University, AlbaNova, SE-106 91 Stockholm, Sweden}
\email{avinash21292@gmail.com}
\author[0009-0000-4068-1320]{Xinlei Chen}
\affiliation{South-Western Institute for Astronomy Research, Yunnan University, Kunming, Yunnan 650500, People's Republic of China}
\affiliation{Yunnan Key Laboratory of Survey Science, Yunnan University, Kunming, Yunnan 650500, People's Republic of China}
\email{xlchen@stu.ynu.edu.cn}
\author[0009-0006-1010-1325]{Yuan~Fang}
\affiliation{South-Western Institute for Astronomy Research, Yunnan University, Kunming, Yunnan 650500, People's Republic of China}
\affiliation{Yunnan Key Laboratory of Survey Science, Yunnan University, Kunming, Yunnan 650500, People's Republic of China}
\email{fangyuan@ynu.edu.cn}
\author[0009-0002-7625-2653]{Yu~Pan}
\affiliation{South-Western Institute for Astronomy Research, Yunnan University, Kunming, Yunnan 650500, People's Republic of China}
\affiliation{Yunnan Key Laboratory of Survey Science, Yunnan University, Kunming, Yunnan 650500, People's Republic of China}
\email{yupan0304@163.com}
\author[0009-0006-5847-9271]{Xingzhu~Zou}
\affiliation{South-Western Institute for Astronomy Research, Yunnan University, Kunming, Yunnan 650500, People's Republic of China}
\affiliation{Yunnan Key Laboratory of Survey Science, Yunnan University, Kunming, Yunnan 650500, People's Republic of China}
\email{xingzhuzou@mail.ynu.edu.cn}
\author[0009-0007-9727-7792]{Hrishav Das}
\affiliation{Indian Institute of Astrophysics, II Block, Koramangala, Bengaluru-560034, Karnataka, India}
\affiliation{Pondicherry University, Chinna Kalapet, Kalapet, Puducherry 605014, India}
\email{hrishav.das@iiap.res.in}
\author[0000-0001-6706-2749]{Mridweeka~Singh}
\affiliation{Indian Institute of Astrophysics, II Block, Koramangala, Bengaluru-560034, Karnataka, India}
\email{yashasvi04@gmail.com}
\author[0009-0008-9062-1455]{Amrit~Dutta}
\affiliation{Indian Institute of Astrophysics, II Block, Koramangala, Bengaluru-560034, Karnataka, India}
\affiliation{Pondicherry University, Chinna Kalapet, Kalapet, Puducherry 605014, India}
\email{amrit.dutta@iiap.res.in}
\author[0000-0003-3533-7183]{G.~C.~Anupama}
\affiliation{Indian Institute of Astrophysics, II Block, Koramangala, Bengaluru-560034, Karnataka, India}
\email{gca@iiap.res.in}
\author[0009-0005-5865-0633]{Xueling~Du}
\affiliation{South-Western Institute for Astronomy Research, Yunnan University, Kunming, Yunnan 650500, People's Republic of China}
\affiliation{Yunnan Key Laboratory of Survey Science, Yunnan University, Kunming, Yunnan 650500, People's Republic of China}
\email{xldu@stu.ynu.edu.cn}
\author[0009-0008-6974-8611]{Yang Yang}
\affiliation{South-Western Institute for Astronomy Research, Yunnan University, Kunming, Yunnan 650500, People's Republic of China}
\affiliation{Yunnan Key Laboratory of Survey Science, Yunnan University, Kunming, Yunnan 650500, People's Republic of China}
\email{yyang@stu.ynu.edu.cn}

\author[0000-0002-6252-3750]{Kaushik Chatterjee}
\affiliation{South-Western Institute for Astronomy Research, Yunnan University, Kunming, Yunnan 650500, People's Republic of China}
\affiliation{Yunnan Key Laboratory of Survey Science, Yunnan University, Kunming, Yunnan 650500, People's Republic of China}
\email{mails.kc.physics@gmail.com}

\author[0009-0005-8762-0871]{Tao Wang}
\email{wangtao@itc.ynu.edu.cn}
\affiliation{South-Western Institute for Astronomy Research, Yunnan University, Kunming, Yunnan 650500, People's Republic of China}
\affiliation{Yunnan Key Laboratory of Survey Science, Yunnan University, Kunming, Yunnan 650500, People's Republic of China}

\author[0009-0003-6936-7548]{Xufeng Zhu}
\email{xufengzhu@ynu.edu.cn}
\affiliation{South-Western Institute for Astronomy Research, Yunnan University, Kunming, Yunnan 650500, People's Republic of China}
\affiliation{Yunnan Key Laboratory of Survey Science, Yunnan University, Kunming, Yunnan 650500, People's Republic of China}

\author[0000-0001-6374-8313]{Yuan-Pei~Yang}
\affiliation{South-Western Institute for Astronomy Research, Yunnan University, Kunming, Yunnan 650500, People's Republic of China}
\affiliation{Yunnan Key Laboratory of Survey Science, Yunnan University, Kunming, Yunnan 650500, People's Republic of China}
\email{ypyang@ynu.edu.cn}
\author[0000-0003-1295-2909]{Xiaowei Liu}
\affiliation{South-Western Institute for Astronomy Research, Yunnan University, Kunming, Yunnan 650500, People's Republic of China}
\affiliation{Yunnan Key Laboratory of Survey Science, Yunnan University, Kunming, Yunnan 650500, People's Republic of China}
\email{x.liu@ynu.edu.cn}
\correspondingauthor{brajesh@ynu.edu.cn, ypyang@ynu.edu.cn}

\begin{abstract}
We present well-sampled multi-band imaging and low- to medium-resolution spectroscopic observations of the early-discovered Type~II supernova (SN)~2025dys. The photometric evolution exhibits a rise time of $\sim$\,18 days ($r$ band) and a rapid decline transitioning to the radioactive decay tail at $\sim$\,70 days. SN~2025dys is among the high-luminosity events with a peak pseudo-bolometric luminosity of $L_{\rm opt}=8.57\times10^{42}\,\mathrm{erg\,s^{-1}}$. Early X-ray emission was identified, displaying a luminosity of $7.55\times10^{40}\,\mathrm{erg\,s^{-1}}$. The observational investigation points to a dense, hydrogen-rich circumstellar medium with a mass of $M_{\rm CSM}\approx0.5$--$0.7\,\mathrm{M_\odot}$, suggesting a partially stripped, hydrogen-poor inner ejecta with a relatively low mass of $M_{\rm ej}\approx2.5\,\mathrm{M_\odot}$. The explosion possessed a kinetic energy of $E_{\rm k}\approx0.3\text{--}0.5$~foe and yielded a synthesized nickel mass of $M_{\rm Ni}\approx0.042_{-0.011}^{+0.012}\,\mathrm{M_\odot}$. These physical parameters indicate an extreme mass-loss episode before the explosion, characterized by a high mass-loss rate of $\dot{M} \gtrsim 0.01\,\mathrm{M_\odot\,yr^{-1}}$. SN~2025dys represents a stripped progenitor, intrinsically resembling a Type~IIb or even a Type~Ib SN, that exploded within a dense, hydrogen-rich circumstellar environment.
\end{abstract}

\section{Introduction}
\label{sec:intro}
Supernovae (SNe) represent the spectacular endpoints of stellar evolution. Classical astrophysical theory delineates two primary physical mechanisms for these extraordinary events: thermonuclear explosions and core collapse (CC). For massive stars ($M_{\rm ZAMS}\gtrsim8$--$10\,\mathrm{M_\odot}$), upon exhausting their nuclear fuel through successive burning stages, the stellar core undergoes gravitational collapse, triggering a core-collapse supernova (CC~SN) \citep{Woosely-1986W, Heger2003H}. Among these events, a subgroup termed stripped-envelope supernovae (SE~SNe) is distinguished by the merely weak presence or complete absence of hydrogen and/or helium features in their optical spectra \citep{1941minkowski, Modjaz2019M}. This spectral deficiency indicates that the progenitor star lost its outer layers prior to the terminal explosion. Such extensive mass loss is typically driven either by intense stellar wind \citep[e.g., from Wolf-Rayet stars or red supergiant superwinds;][]{Smith2014ARAA, Puls-2008P, Chiosi-1986C} or via mass transfer to a binary companion through Roche-lobe overflow (RLOF) \citep[e.g.,][]{Podsiadlowski-1992P, Yoon-2010Y}.

Among these diverse transients, certain peculiar events exhibit remarkably high luminosities. This enhanced emission can be powered either by a central engine, such as a rapidly spinning magnetar \citep{Omand2024MNRAS.527.6455O}, or through violent interaction with dense circumstellar material (CSM) \citep{Chatzopoulos2012apj, Dessart2018AA}. For luminous Type~II supernovae (SNe~II), CSM interaction is the most prevalent mechanism, providing kinetic energy dissipation that can far exceed the energy deposition from radioactive decay. Evidence of such interaction often appears in the form of early-phase flash-ionization signatures \citep{Gal-Yam-2014G}. The first well-studied event with these features was SN~1998S \citep{Leonard-2000L, SN1998S_Anupama2001A&A...367..506A}. The two nearby SNe 2023ixf and 2024ggi have also shown similar evidence \citep[e.g.][]{Jacobson-2023J, Gaici-Li2024L, Shrestha-2024S, Jacobson-2024J}. These events exhibit enhanced luminosity during the first few days before transitioning to a normal SN~II evolution. Conversely, strongly interacting events like SN~2006gy and SN~2010jl maintain persistent CSM interaction over extended periods, implying a massive, widespread circumstellar environment \citep[e.g.][]{Ofek2007ApJ...659L..13O, Moriya2020A&A...641A.148M}. 

Recently, a rare and intriguing interacting scenario has emerged involving a stripped-envelope progenitor (e.g., a bare helium star) exploding within a dense, hydrogen-rich CSM. This configuration distinctly contrasts with typical H-rich SNe~IIn \citep[e.g.,][]{SN1998S_Anupama2001A&A...367..506A, Ofek2007ApJ...659L..13O} or He-rich SNe~Ibn \citep[e.g.,][]{SN2006jc_Anupama2009MNRAS.392..894A, Cai2026A&A...707A.157C}. Well-known examples of this phenomenon include SN~2014C and SN~2019oys \citep{SN2014C_Milisavljevic2015ApJ...815..120M, Sollerman2020A&A...643A..79S}. These transients initially appeared as ordinary hydrogen-poor SNe~Ib but later metamorphosed into interacting SNe~IIn as their expanding ejecta crashed into a distant, hydrogen-rich shell. In these events, the observations point to a fully detached H-rich CSM shell, where the interaction primarily occurs during the post-peak or late-time phases, producing prominent bumps in the light curve evolution. The formation of such distant shells can be explained by extreme RLOF episodes within a binary progenitor system \citep{SN2014C_2025ApJ...978..163Z}. Motivated by these scenarios to find the transitional type of these events, we present the observations and analysis of the early-discovered SN~2025dys. This event filled the observational gap for this class of transients. As we will demonstrate in this article, SN~2025dys represents a stripped helium star exploding within an asymmetric, confined, and incompletely detached dense H-rich CSM, providing unprecedented insight into the immediate pre-supernova mass-loss phase.

\section{Discovery and follow-up}\label{sec:disc}

\begin{figure}%[htbp]
\centering
\includegraphics[width=\linewidth]{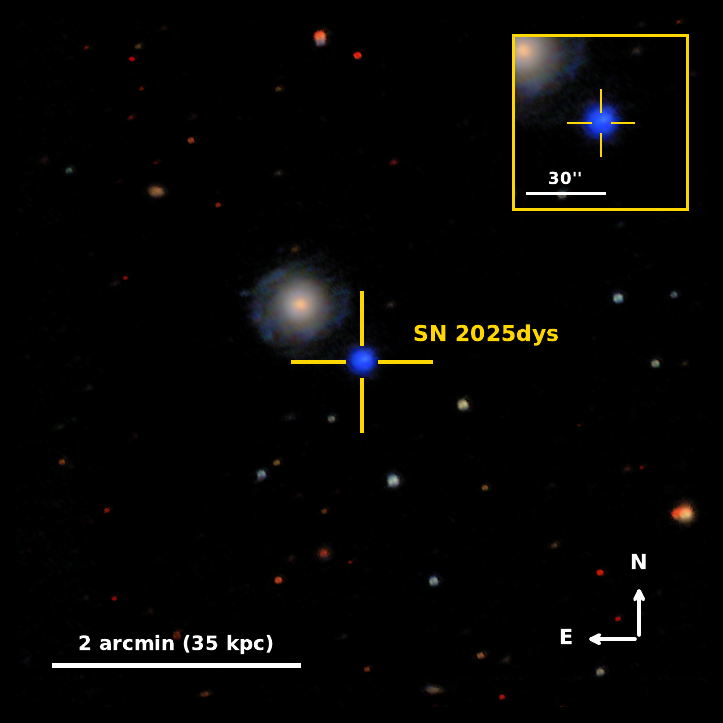}
\caption{SN~2025dys with its host LEDA~3081781 taken from Mephisto BigCamera. The colored image was synthesized by stacking the $u_{\rm M}v_{\rm M}g_{\rm M}r_{\rm M}i_{\rm M}z_{\rm M}$ bands with each corresponding zero-point.}
\label{fig:misc:fc}
\end{figure}

SN~2025dys ($\mathrm{RA}=10^{\mathrm{h}}50^{\mathrm{m}}19\fs89$, $\mathrm{DEC}=-22\arcdeg19\arcmin49\farcs37$) was discovered by the Asteroid Terrestrial-impact Last Alert System (ATLAS) \citep{Tonry2018apj, Tonry2018pasp} on 2025 March 09 (MJD~60743.93), with an apparent magnitude of $18.873\pm0.168$~mag in the ATLAS orange ($o$) band \citep{Tonry2025TNSTR.938....1T}. The last non-detection of SN~2025dys was taken on March 8$^{\rm th}$, 2025 (MJD~60742.25), with a limiting magnitude of $19.36$~mag. The field of SN~2025dys is shown in Fig.~\ref{fig:misc:fc}.  The most possible host galaxy of SN~2025dys is LEDA~3081781, which is $31\farcs2$ north and $26\farcs7$ east of SN~2025dys. SN~2025dys was classified as a near-maximum SN~II on March $17^{\rm th}$ 2025 \citep{Stritzinger2025TNSCR1035....1S}. 

\begin{figure*}%[htbp]
\centering
\includegraphics[width=\linewidth]{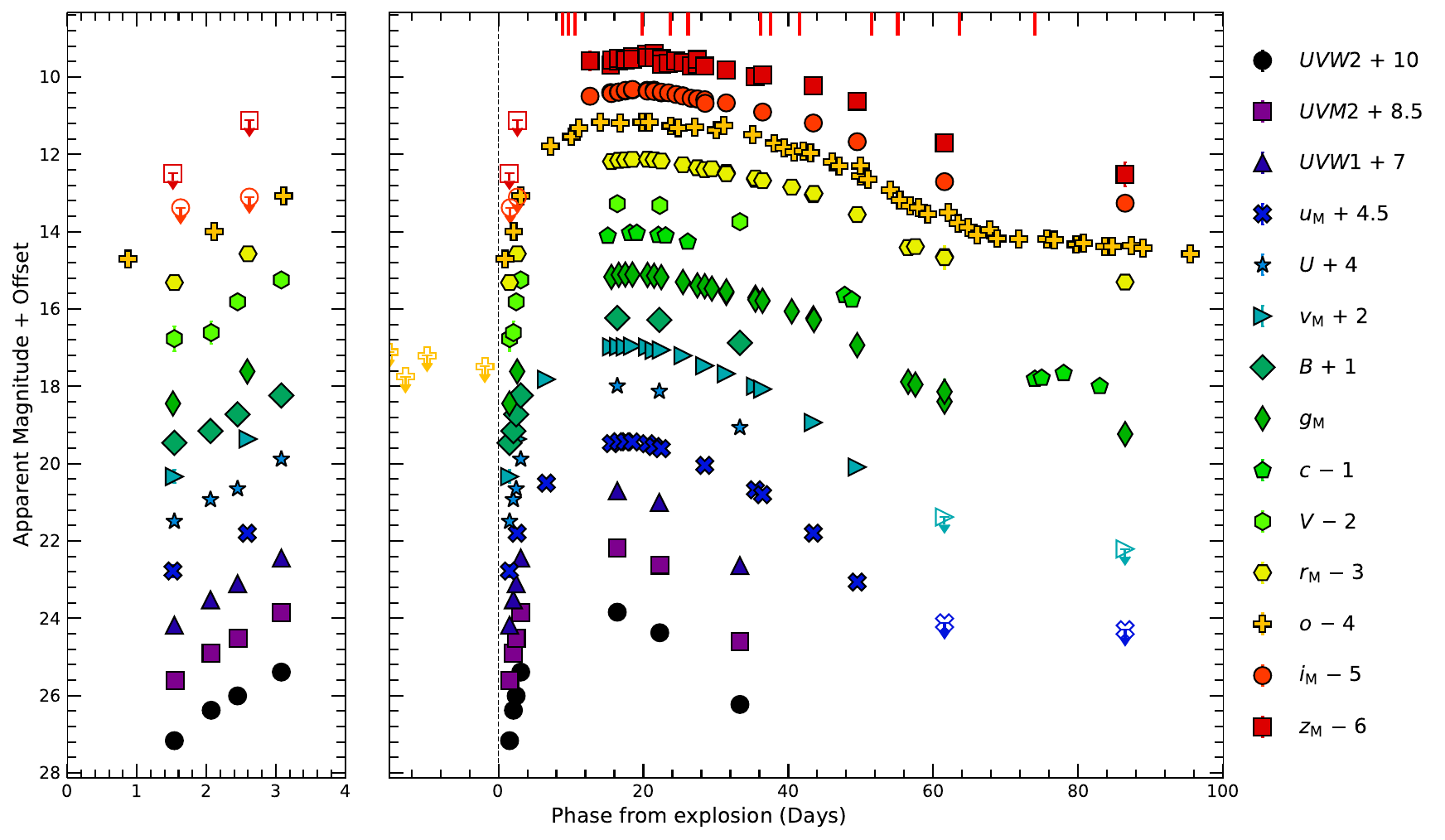}
\caption{Multi-band UV and optical apparent light curve of SN~2025dys. The dashed vertical line refers to the estimated explosion epoch (see Section~\ref{sec:phot}.) The short red lines vertically at the top mark the epochs of the spectroscopic observations. Upper limits are plotted with hollow symbols with arrows. The light curves for different filters are shifted correspondingly with different colours.}
\label{fig:phot:applc}
\end{figure*}

We considered the redshift measurement provided by \cite{Palau2025TNSCR1202....1P} and took the uncertainty from the inconsistency between the different classification reports, thus yielding a result of $z=0.014\pm0.001$. We take the assumption of a flat $\Lambda \mathrm{CDM}$ cosmology with $H_{0}=73\,\mathrm{km\,s^{-1}\,Mpc^{-1}}$, $\Omega_{\rm M}=0.27$ and $\Omega_{\Lambda}=0.73$, yield a luminosity distance of $D_{\rm L}=58.1\pm4.2\,\mathrm{Mpc}$ \citep{Spergel2007apjs}. The corresponding distance modulus is $\mu=33.82 \pm 0.16$~mag. The total extinction in the direction of SN 2025dys is a combination of foreground Galactic extinction and extinction within the host galaxy. The foreground extinction for SN~2025dys is $E(B-V)_{\mathrm{MW}}=0.055\pm0.003$~mag \citep{2011ApJ...737..103S}. We cannot identify the \ion{Na}{i}~D line in the early spectrum of SN~2025dys; thus, the extinction from the host galaxy could be negligible. Thus, the total line-of-sight extinction $E(B-V)_{\rm tot}=0.055\pm0.003$ has been considered in our analysis. 

The global properties of LEDA~3081781 were investigated with \textsc{prospector} code \citep{Johnson2021ApJS..254...22J, Conroy2009ApJ...699..486C}. It infers $Z \approx 0.17\,\mathrm{Z_\odot}$, a significantly sub-solar metallicity. Further, a total stellar mass of $M_{*}\approx10^{10}\,\mathrm{M_\odot}$ was estimated, indicating this host to be a moderately massive system (less massive than the Milky Way). A detailed investigation of the star formation history and its possible influence on the progenitor of SN~2025dys can be found in Appendix~\ref{sec:app:host}.

The imaging of SN~2025dys was promptly initiated from 2025 March 10 (MJD~60744.61) using 50\,cm array and 1.6\,m Multi-channel Photometric Survey Telescope (Mephisto) \citep[see,][]{Zhang-2025ATel}. Simultaneous multi-band frames were acquired in Mephisto $u_{\rm M}v_{\rm M}g_{\rm M}r_{\rm M}i_{\rm M}z_{\rm M}$ bands \citep[for filter specifications, see][]{Yang2024ApJ...969..126Y, Chen2024ApJ...971L...2C, SN2025aico_Zhao2026arXiv260710671Z}. The photometric data were supplemented from the Asteroid Terrestrial-impact Last Alert System \citep[ATLAS,][]{Tonry2018pasp} in $cyan$ and $orange$ bands, and \textit{Swift}/UVOT \citep{Gehrels2004G} in  $UVW2, UVM2, UVW1, U, B$ and $V$ bands. Further, \textit{Swift}/XRT data were also utilized. The spectroscopic investigation was performed using a total of 13 spectra acquired at the 2\,m Himalayan Chandra Telescope (HCT) and the 3.6\,m ESO New Technology Telescope (NTT) public archive\footnote{PI: Inserra; Program ID: 112.25JQ} and also from the Transient Name Server archive. Details on data reduction can be found in Section~\ref{sec:app:ctf:red}.

\section{Photometry}\label{sec:phot}
\subsection{Light curve evolution}\label{sec:phot:lc}

The early-phase ultraviolet (UV) and optical ($uvgriz$) light curves of SN~2025dys are presented in Fig.~\ref{fig:phot:applc}. The photometric observations provide excellent coverage starting from the discovery epoch and extending to approximately $100$~days post-discovery. The \textit{Swift}-UVOT photometric data are reported in the Vega magnitude system, whereas all other measurements are presented in the AB magnitude system. The $u_{\rm M}$ through $z_{\rm M}$ bands represent the dedicated filter set of the Mephisto telescope, providing optimized bandpasses in the near-UV and blue regimes, alongside extended coverage at redder wavelengths. A comprehensive description of the Mephisto $u_{\rm M}v_{\rm M}g_{\rm M}r_{\rm M}i_{\rm M}z_{\rm M}$ system can be found in \cite{Yang2024ApJ...969..126Y, Chen2024ApJ...971L...2C, SN2025aico_Zhao2026arXiv260710671Z}. To estimate the explosion epoch, we utilized Gaussian Process (GP)\footnote{See \url{https://scikit-learn.org/stable/modules/gaussian_process.html}} regression to extrapolate the early light curve. Following the method described by \cite{SN2017ati_Peng2026A&A...709A.156P}, the explosion time and associated uncertainty are determined from the epoch at which the extrapolated flux reaches zero. Specifically, we applied this extrapolation to the ATLAS $o$-band data, as these observations offer optimal early-phase coverage. This approach yields an explosion epoch of $t_0 = 60743.05 \pm 0.19$ (in the MJD frame). As a consistency check, we calculated the midpoint between the last non-detection and the first detection in the ATLAS $o$-band. This calculation results in an epoch of MJD~$60743.09 \pm 0.86$, which is fully consistent with the GP estimation. Therefore, we adopted the GP-derived result as the final explosion epoch due to the significantly lower uncertainty.

We applied polynomial fitting to the light curve of SN~2025dys and utilized a Monte-Carlo simulation to estimate the uncertainties of the fitting parameters. We determined the peak luminosity epoch to be MJD~$60761.31_{-0.69}^{+0.51}$ in the $g_{\rm M}$-band, and MJD~$60761.41_{-0.86}^{+0.65}$ in the $r_{\rm M}$-band. The corresponding peak magnitudes are $15.176_{-0.003}^{+0.002}$~mag in the $g_{\rm M}$-band and $15.133\pm0.011$~mag in the $r_{\rm M}$-band. These values indicate a rise time of $t_{\rm rise}\approx18$~days, pointing to a longer rise duration compared with the majority of SNe~II events \citep{Anderson2014ApJ...786...67A, Valenti2016MNRAS.459.3939V}. After the peak magnitude, the decline of SN~2025dys can be described by three linear segments. We applied linear fits to both the $g_{\rm M}$ and $r_{\rm M}$ bands, revealing a turning point in the decline rate at $\sim20$~days post-peak, located between the first and second linear phases. The initial decline rate is $\gamma_{t\lesssim20}=3.43\pm0.19\,\mathrm{mag\,(100d)^{-1}}$ in the $g_{\rm M}$-band and $\gamma_{t\lesssim20}=3.54\pm0.23\,\mathrm{mag\,(100d)^{-1}}$ in the $r_{\rm M}$-band. Following the turning point, the decline rate increases to $\gamma_{20\lesssim t\lesssim60}=11.43\pm0.47\,\mathrm{mag\,(100d)^{-1}}$ in the $g_{\rm M}$-band and $\gamma_{20\lesssim t\lesssim60}=9.17\pm0.41\,\mathrm{mag\,(100d)^{-1}}$ in the $r_{\rm M}$-band. This phenomenon, where the bluer bands drop faster than the redder bands, might be caused by the late-phase cooling of the supernova ejecta.

Because of the limited coverage during the late phase in the Mephisto $u_{\rm M}v_{\rm M}g_{\rm M}r_{\rm M}i_{\rm M}z_{\rm M}$ bands, we utilized the ATLAS $o$ band to analyse the late-time light curve evolution. We observed that during the late phase, the $o$-band light curve presents a second break point approximately $70$~days after the estimated explosion epoch, indicating that the luminosity after this point is primarily powered by radioactive decay. This duration is shorter than the typical duration observed in most Type II supernovae with early interaction (including SNe~II-P and SNe~II-L), which usually last around $100$~days. The decline rate after the second break point is $2.14 \pm 0.09\,\mathrm{mag\,(100d)^{-1}}$. This decline is significantly faster than the rate expected from fully trapped $^{56}\mathrm{Co}\rightarrow^{56}\mathrm{Fe}$ decay \citep[approximately $0.98\,\mathrm{mag\,(100d)^{-1}}$;][]{Anderson2014ApJ...786...67A, Valenti2016MNRAS.459.3939V}, a discrepancy that can be attributed to $\gamma$-ray leakage, which is common in those of SE-SNe. If we consider SN~2025dys as a hydrogen-rich event, the plateau duration (if present) must be bounded by $t_{\rm PT}\lesssim70$~days. This characteristic resembles typical short-plateau scenarios \citep[such as SN~2020jfo and SN~2018gj; see][]{SN2018gj_Teja2023ApJ...954..155T, SN2020jfo_Teja2022ApJ...930...34T}. Alternatively, the transient could represent a partially stripped event, similar to SNe~IIb or SNe~Ib, that exploded within a hydrogen-rich CSM. Such a partially stripped envelope naturally accounts for the enhanced $\gamma$-ray leakage.

\begin{figure*}[htbp]
\centering
\includegraphics[width=\linewidth]{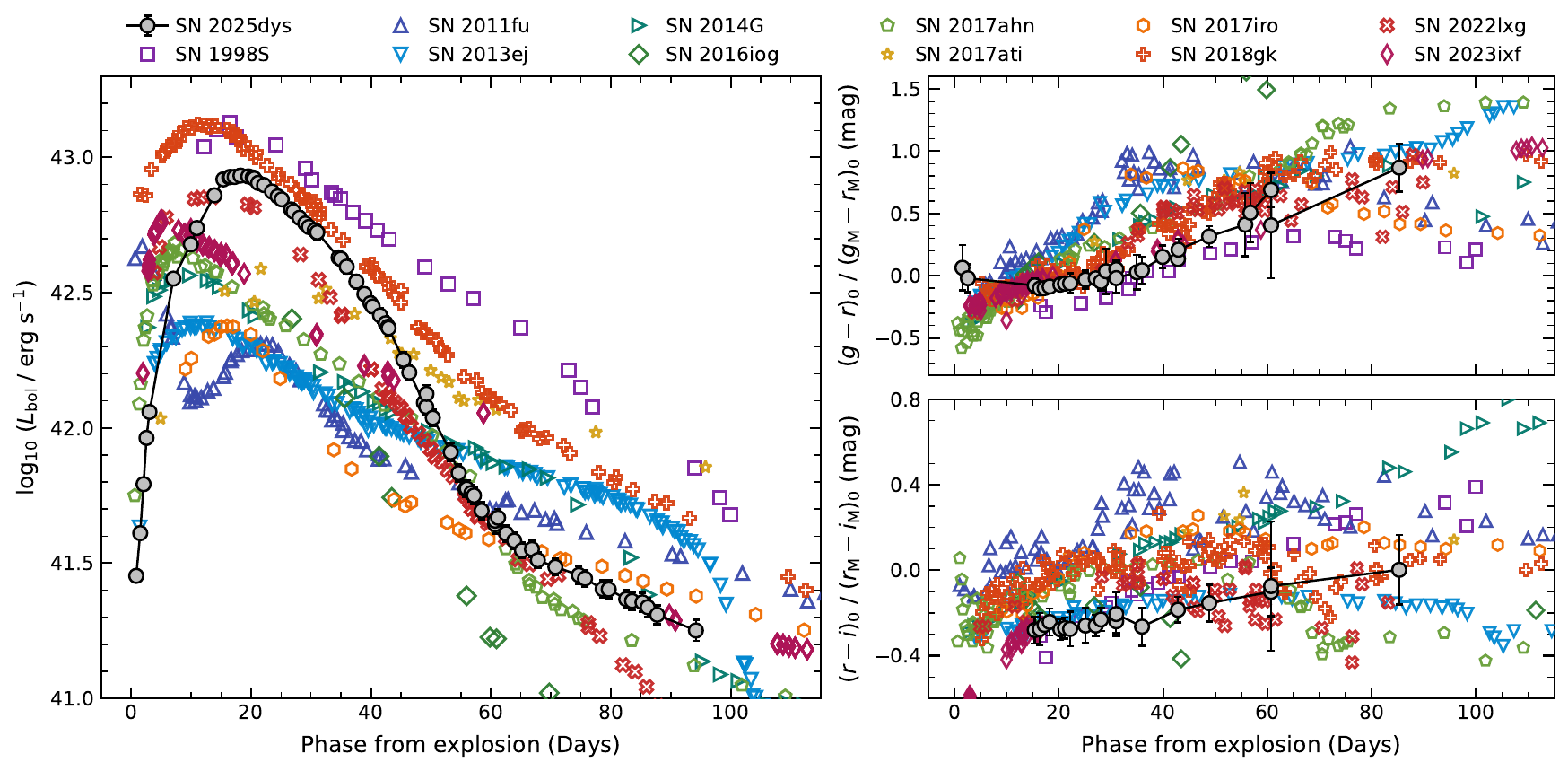}
\includegraphics[width=\linewidth]{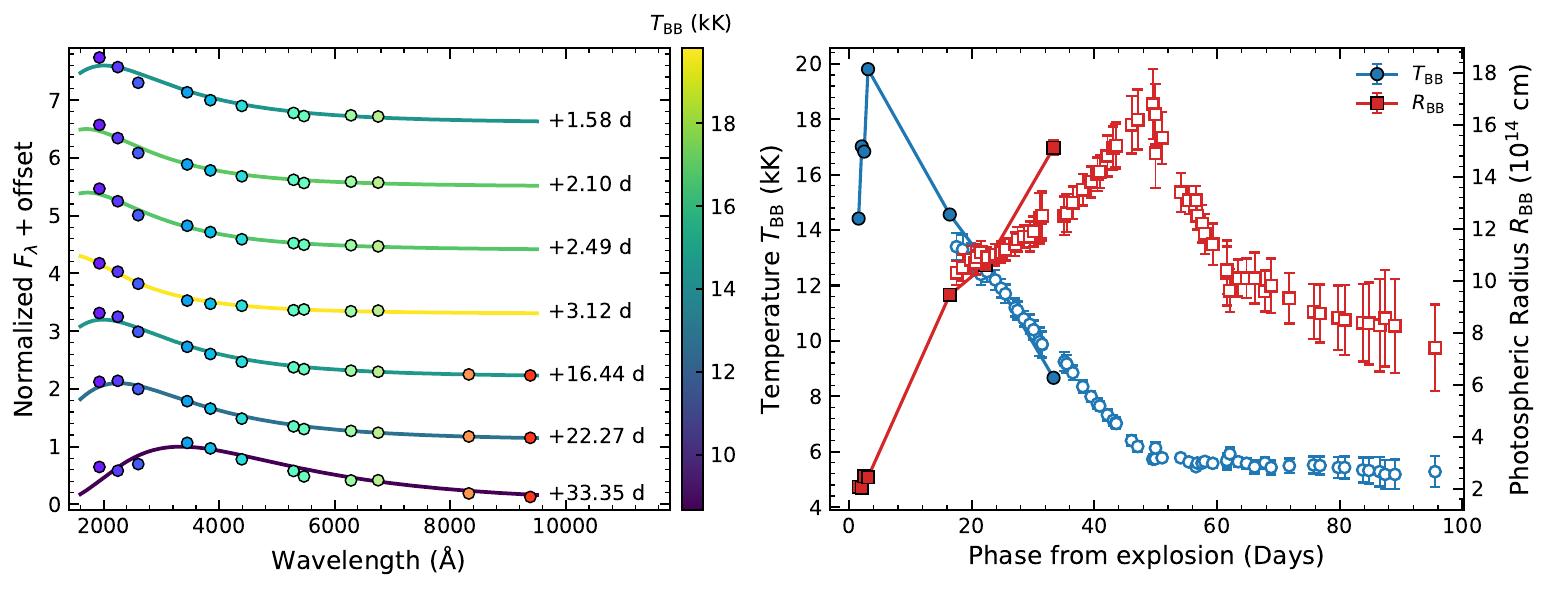}
\caption{\textit{Upper left panel}: Optical pseudo-bolometric light curve of SN~2025dys compared with a reference sample. \textit{Upper right panel}: colour evolution of SN~2025dys and comparison SNe. Note that we approximate the standard $g-r$ and $r-i$ colours using $g_{\rm M}-r_{\rm M}$ and $r_{\rm M} - i_{\rm M}$, respectively. For comparison targets with only $UBVRI$ photometry, the colours are converted from $B-V$ and $R-I$. \textit{Lower left panel}: Early-phase UV and optical spectral energy distribution (SED) fitting of SN~2025dys, where the colour of the SED curves indicates the derived blackbody temperature. \textit{Lower right panel}: Evolution of the blackbody temperature and photospheric radius of SN~2025dys. Solid points denote the values derived from the SED fitting as shown in the left panel, while open markers represent the results obtained using the \textsc{superbol} code with proper interpolation.}
\label{fig:phot:bolcolor}
\end{figure*}

We construct the pseudo-bolometric light curve of SN~2025dys using the \textsc{superbol} code \citep{Nicholl2018RNAAS}. The luminosity is calculated by integrating the fluxes across the optical bands ($u_{\rm M}v_{\rm M}g_{\rm M}r_{\rm M}i_{\rm M}z_{\rm M}$ and the ATLAS $c$ and $o$ bands). We apply reddening corrections to the data, and the results are presented in Fig.~\ref{fig:phot:bolcolor}. We observe that SN~2025dys is a highly luminous transient, with a peak luminosity higher than most objects in our comparison sample, including typical H-poor SNe~IIb and SNe~Ib, as well as some early interacting events. The peak luminosity of SN~2025dys reaches $L = (8.57 \pm 0.41) \times 10^{42}\,\mathrm{erg\,s^{-1}}$. This value exceeds the typical peak luminosities of the comparison sample, which generally range from $\approx 4.0 \times 10^{42}$ to $5.0 \times 10^{42}\,\mathrm{erg\,s^{-1}}$. Instead, it is comparable to the highly luminous SN~1998S (approximately $10^{43}\,\mathrm{erg\,s^{-1}}$) and SN~2022lxg. This suggests that SN~2025dys is unlikely to be powered solely by radioactive decay but by an additional energy source. Thus, in this work, we suggest that SN~2025dys likely experiences relatively strong CSM interaction, similar to the interaction observed in SN~1998S. An alternative explanation could be the central engine-powered scenario like SN~2018gk, which shared a similar peak luminosity to SN~2025dys. However, we are less in favour of this explanation because, during the decline phase, SN~2025dys exhibits a series of segmented declines and a more rapid evolution compared to magnetar-driven luminous events like SN~2017ati and SN~2018gk. This behavior suggests that the high luminosity of SN~2025dys is less likely to be produced by a central engine \citep{SN2017ati_Peng2026A&A...709A.156P, Wang2023ApJ...948..138W, SN2018gk_Bose2021MNRAS.503.3472B}. Moreover, spectroscopic analysis also shows evidence of strong interaction, which we discuss later.

The pseudo-bolometric light curve of SN~2025dys has a relatively long rise time compared to the rest of the sample, with $t_{\rm rise} = 19.48 \pm 0.81$~days. In contrast, events with strong early-time interaction, such as SN~2023ixf and SN~2017ahn, typically exhibit faster rise times of about $5$ to $10$~days \citep{Yang2024ApJ...969..126Y, Tartaglia2021ApJ...907...52T, Dastidar2024A&A...685A..44D}. Such a prolonged rise time indicates a longer photon diffusion timescale, which suggests either a lower kinetic energy or a massive CSM. However, the post-peak fast evolution of SN~2025dys is remarkable. It evolves faster than most of our comparison objects, although slower than extreme events like SN~2016iog \citep{Peng2026A&A...705A.104P}. It does not exhibit a typical long linear decline or a plateau phase, unlike SN~2013ej and SN~2014G \citep{Bose2015ApJ...806..160B, Bose2016MNRAS.455.2712B}. This lack of a prolonged plateau suggests that the dense CSM around SN~2025dys does not possess an extended radial profile. In addition, we note that the late-time tail of SN~2025dys is similar to the comparison objects that have lower peak luminosities. Also, the light curve decline rate becomes slower, which resembles those targets dominated by radioactive power like SN~2011fu and SN~2017iro \citep{Kumar2013mnras, SN2017iro_Kumar2022ApJ...927...61K}. This confirms that the early-time light curve of SN~2025dys must have an extra power input.

The colour evolution of SN~2025dys is generally bluer than that of most reference events, suggesting that the interaction region maintains a persistently high temperature. Taking the intrinsic $(g_{\rm M}-r_{\rm M})_0$ colour into account, we observe that during the earliest epoch (from $+1.5$ to $+2.5$~days after explosion), the optical emission is relatively red, with a value of $0.044\pm0.112$~mag. This characteristic is further supported by the blackbody temperature evolution derived from spectral energy distribution (SED) fitting, as shown in the lower-left panel of Fig.~\ref{fig:phot:bolcolor}. At $+1.58$~days, the derived temperature is approximately $14,000$~K, which subsequently rises to a peak of around $20,000$~K by $+3.12$~days after explosion. This initial heating phase can be attributed to a delayed photon diffusion process operating within a dense, optically thick circumstellar medium, or the delayed arrival of the forward shock at the inner boundary of the dense circumstellar shell at $t \approx 3.12$~days. Alternatively, from an observational perspective, because the H$\alpha$ line falls directly within the passband of the $r_{\rm M}$ filter, the presence of strong early-time H$\alpha$ emission could artificially redden the measured colour. Thus, line blending provides a third plausible explanation for this initial red phase.

At approximately $15$~days after the explosion, the colour evolution starts to follow the typical trend of the comparison sample. The $(g_{\rm M}-r_{\rm M})_0$ colour increases linearly from $-0.014\pm0.054$~mag to $0.932\pm0.193$~mag between $+15$ and $+85$~days. We observe that the $(r_{\rm M}-i_{\rm M})_0$ colour follows a similar trajectory, rising from $-0.280\pm0.062$~mag to $0.001\pm0.162$~mag over the same period. The only discrepancy arises in the shape of the colour evolution profile. For targets such as SN~2017ahn and SN~1998S, the colour evolution exhibits a double-linear profile with a distinct break point, indicating the transition from the optically thick phase to the nebular phase \citep{Tartaglia2021ApJ...907...52T}. However, we cannot identify a similar break point in the colour evolution of SN~2025dys. Combined with the results from our blackbody temperature fitting, the transition starts at approximately $+50$~days, when the temperature reaches the hydrogen recombination threshold and the photosphere begins to recede. Despite this physical transition, a corresponding break is not reflected in the colour curves. This absence might be caused by specific flux contributions from emission lines. For SN~2025dys, there is persistent interaction with a confined CSM, alongside strong intrinsic line emission (such as the \ion{He}{i}~$\lambda6678$ line) from the ejecta of the partially stripped progenitor. These lines contribute significantly more flux to the $r_{\rm M}$ band than to other bands, dominating the broadband flux and thereby masking the expected continuum transition in the colour evolution.

\subsection{Light curve modeling}\label{sec:phot:lcmdl}

The high-luminosity light curve of SN~2025dys likely results from the combined contributions of radioactive decay (RD) and CSM interaction (CSI). Specifically, CSI drives the early high-luminosity, fast-evolving peak, while RD powers the subsequent low-luminosity, linearly declining tail (in magnitude space). Consequently, we adopt the theoretical framework developed by \cite{Chatzopoulos2012apj}, which combines the RD and CSI mechanisms with photon diffusion. To implement this, we employ the \textsc{mosfit} package \citep{Guillochon2018ApJS..236....6G} using the \textsc{csmni} model. Parameter space sampling is performed via the nested sampling library \textsc{dynesty} \citep{dynesty_Speagle2020MNRAS.493.3132S}. This modeling approach is widely used to analyse interacting transients, such as SNe~IIn and SNe~Ibn \citep[e.g.,][]{Rameshan_2026R, Wang2025A&A...700A.156W}. In addition, for the sake of completeness, we implement a central engine scenario driven by magnetar spin-down, a mechanism that can also produce high luminosity \citep[e.g.,][]{Maeda2007ApJ...666.1069M, Omand2024MNRAS.527.6455O}. As a control for comparison, we include a baseline model powered solely by radioactive decay. The corresponding \textsc{mosfit} models for these two alternative scenarios are \textsc{magni} and \textsc{default}, respectively. For the multi-band fitting, we incorporate photometric data from the Mephisto $uvgriz_{\rm M}$ bands, the \textit{Swift} ultraviolet bands, and the ATLAS $c$ and $o$ bands. Fitting the multi-band photometry simultaneously constrains both the luminosity across different bands and the overall colour evolution, effectively breaking degeneracies among physical parameters.

\begin{figure*}[htbp]
\centering
\gridline{\fig{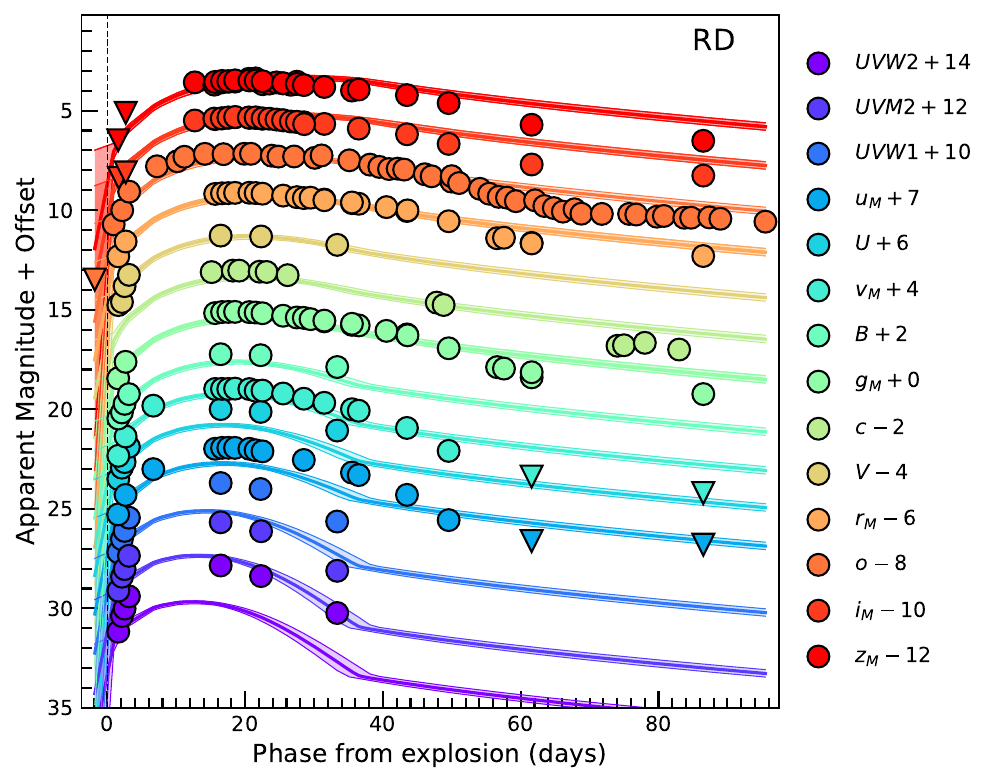}{0.5\textwidth}{}
          \fig{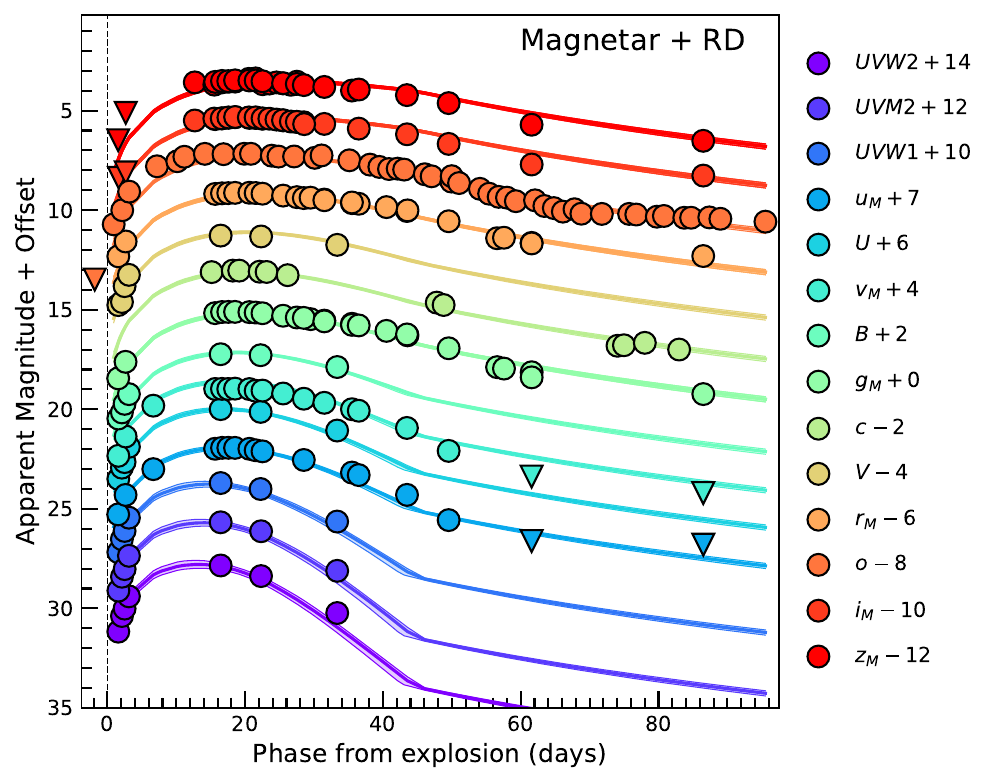}{0.5\textwidth}{}}
\gridline{\fig{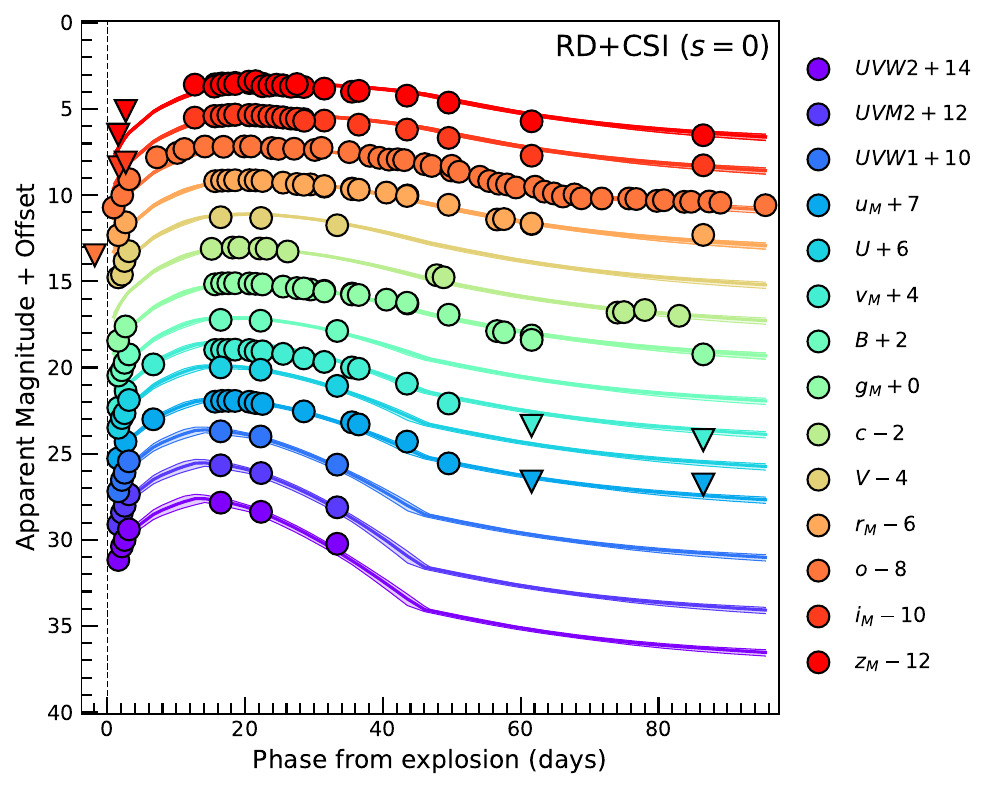}{0.5\textwidth}{}
          \fig{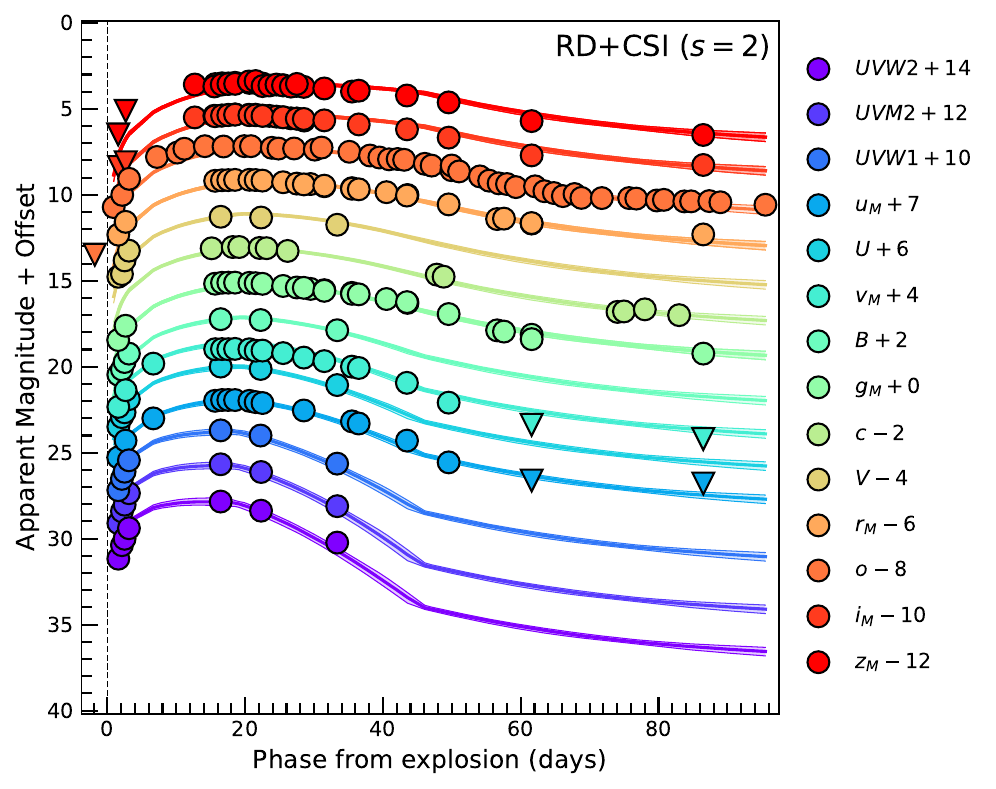}{0.5\textwidth}{}}
\caption{Multi-band light curve fitting result of SN~2025dys. The observed detections are marked with circle dots, while the upper-limit data are marked with triangle points. The best-fit results of each band are marked with solid lines, and the shaded region represents the $\pm1\sigma$ confidence interval. Different bands are presented in different colours. Different models are marked as text in the upper right corners of each plot.}
\label{fig:phot:lcmdl}
\end{figure*}

The \textsc{csmni} model is governed by several physical parameters. The photon diffusion process depends on the optical opacity ($\kappa_{\rm opt}$), the ejecta mass ($M_{\rm ej}$), the kinetic energy ($E_{\rm K}$, which corresponds to the ejecta velocity), and the geometry of the ejecta. Given that the diffusion region is hydrogen-rich, we adopt an opacity of $\kappa_{\rm opt}=0.34\,\mathrm{cm^{2}\,g^{-1}}$, which is a typical value for hydrogen-rich environments \citep[e.g.,][]{Nagy2014AA, Nagy2018ApJ, Rameshan_2026R}. Because most of such transients originate from extended progenitors, we apply a broken power-law density profile for the ejecta. We fix the inner density index to $\delta=0$ and the outer density index to $n=12$, which accurately represents the structure of an extended progenitor, such as a yellow or red supergiant \citep[see, e.g.,][]{Charalampopoulos2025A&A...700A.138C}. The CSI process is characterized by the CSM mass ($M_{\rm CSM}$), the inner radius of the CSM ($R_0$), and the CSM density distribution. We assume a simple power-law radial density profile, $\rho\propto r^{-s}$, which is controlled by the density slope ($s$) and a reference density ($\rho_0$). To ensure robust parameter estimation, we test two different CSM density profiles to fit the data of SN~2025dys: $s=0$ (representing a shell-like CSM) and $s=2$ (representing a steady stellar wind). For the magnetar-driven scenario, the model is primarily governed by the magnetic field strength ($B$) and the initial spin period ($P$). We fix the mass of the central neutron star to $M_{\rm NS}=1.4\,\mathrm{M_\odot}$ and ignore the effect of the viewing angle, as it remains poorly constrained by light curve fitting. In addition, a minimum photospheric temperature ($T_{\rm min}$) is introduced to regulate the late-time photospheric evolution and generate the multi-band light curves; however, since the final fitting results are generally insensitive to this specific parameter \citep{Wang2025A&A...700A.156W}, we fix the temperature by adopting the floor blackbody temperature present in Sect.~\ref{sec:phot:lc}, which is $5700$~K\footnote{See the bottom right figure in Fig.~\ref{fig:phot:bolcolor}.}. The radioactive decay chain follows the $^{56}{\rm Ni}\rightarrow ^{56}{\rm Co}\rightarrow ^{56}{\rm Fe}$ transition, which is parametrized by the nickel fraction ($f_{\rm Ni}$). To model any underestimated systematic uncertainties in the data, we incorporate a variance parameter ($\log \sigma$) as an extra jitter term. Ultimately, we perform the fitting by utilizing $120$ live points and allowing the sampler to run until it fully meets the convergence criteria\footnote{The models are considered to have good convergence until $\mathrm{d}\log z\lesssim 0.01$.}.

The fitting results for all previously discussed models are presented in Fig.~\ref{fig:phot:lcmdl}. We note that the pure RD model cannot accurately reproduce the light curves of SN~2025dys, exhibiting a significant discrepancy, particularly in the NUV bands. Furthermore, the inferred nickel mass fraction in the RD model approaches unity, which is physically unrealistic. The typical nickel mass ranges from $0.01$ to $0.1\,\mathrm{M_\odot}$ for SNe~II and from $0.1$ to $0.2\,\mathrm{M_\odot}$ for SE-SNe \citep{Taddia2018A&A, Anderson2014ApJ...786...67A, Valenti2016MNRAS.459.3939V}, both of which are substantially lower than the predictions of this model. This suggests that an additional energy source is required to power SN~2025dys. For the CSI scenarios, both the $s=0$ and $s=2$ models successfully reproduce the light curves of SN~2025dys. They yield similar posterior distributions for most parameters, which are presented in Appendix~\ref{sec:app:ctf:mcmcpost}. This indicates that the progenitor of SN~2025dys likely underwent an eruptive mass-loss process prior to explosion, forming a confined, shell-like CSM. This conclusion is consistent with our prior expectations. 

The best-fit models suggest a low-to-intermediate energy explosion. Specifically, the $s=0$ model yields an ejecta mass of $M_{\rm ej}=2.45_{-0.37}^{+0.30}\,\mathrm{M_{\odot}}$ and a kinetic energy of $E_{\rm K}=0.37\pm0.06\,\mathrm{foe}$, while the $s=2$ model yields $M_{\rm ej}=2.69^{+0.33}_{-0.35}\,\mathrm{M_{\odot}}$ and $E_{\rm K}=0.46_{-0.06}^{+0.04}\,\mathrm{foe}$. Because these values of $M_{\rm ej}$ are smaller than those typically observed for SNe~II-L or SNe~II-P, the progenitor is likely a partially stripped star, bearing similarities to the progenitors of SNe~IIb or even SNe~Ib \citep[e.g.][]{Nagy2016aap, SN2022ngb_Zhao2026A&A}. The corresponding ejecta velocity is $v_{\rm ej}\approx5200\,\mathrm{km\,s^{-1}}$ for both models. This velocity is lower than the typical expansion velocities of stripped-envelope transients, such as SNe~Ibc and SNe~IIb, which generally range from $8000$ to $10000\,\mathrm{km\,s^{-1}}$ near peak luminosity \citep[see][]{Holmbo2023A&A}, but similar to those SE-SNe with lower energy like SN~2025aico \citep{SN2025aico_Zhao2026arXiv260710671Z}. 

The inferred CSM masses for SN~2025dys are $M_{\rm CSM}=0.76_{-0.10}^{+0.13}\,\mathrm{M_\odot}$ and $0.54_{-0.02}^{+0.05}\,\mathrm{M_\odot}$ for the $s=0$ and $s=2$ models, respectively. These values are significantly more massive than the CSM in similar events like SN~2022lxg and SN~2016iog, where the CSM mass is generally $\lesssim0.1\,\mathrm{M_\odot}$ \citep{Peng2026A&A...705A.104P, Charalampopoulos2025A&A...700A.138C}. The most significant discrepancy between the two models arises in the estimated inner radius. For the $s=0$ model, the inner radius is estimated to be $R_0=83_{-12}^{+17}\,\mathrm{AU}$ with a uniform density of $\rho_0=1.82_{-0.41}^{+0.53}\times10^{-12}\,\mathrm{g\,cm^{-3}}$. In contrast, the $s=2$ model yields an inner radius of $R_0=6.2_{-4.8}^{+9.7}\,\mathrm{AU}$ with a density of $\rho_0=1.17_{-0.97}^{+23}\times10^{-11}\,\mathrm{g\,cm^{-3}}$. We note that the parameters produced by the $s=0$ model are well-constrained, while the density in the $s=2$ model exhibits a large uncertainty. Furthermore, the inner radius derived from the $s=0$ model is extremely extended, which is considered highly unphysical. This feature is also present in \cite{SN2020able_Tartaglia2026arXiv260807661T}, suggesting that certain parameters derived from \textsc{mosfit} should be treated as mathematical fits. If we consider the inner radius resulting from the $s=2$ case, the corresponding outer radius of the CSM is $R_{\rm out}\approx63\,\mathrm{AU}$ (approximately $10^{15}\,\mathrm{cm}$), resembling a dense and confined CSM environment. The best-fit nickel mass of both $s=0$ and $s=2$ models are $M_{\rm Ni}\approx0.045\,\mathrm{M_\odot}$. This value follows most of those of SNe~II, and some of the lower luminosity SE-SNe like SN~2022ngb and SN~2025aico \citep{SN2025aico_Zhao2026arXiv260710671Z, SN2022ngb_Zhao2026A&A}. We noticed that, regardless of the difference in the inner radius $R_0$, the best-fit results for both $s=0$ and $s=2$ are similar, which present a relatively low ejecta mass of $\sim2.5\,\mathrm{M_\odot}$ with a confined CSM with mass $\sim0.5\text{--}0.7\,\mathrm{M_\odot}$. 

In addition, we test the magnetar-driven model combined with radioactive decay. This model also provides a good fit to the data. The best-fit parameters suggest a magnetar central engine with a magnetic field of $B\approx4.7\times10^{14}\,\mathrm{G}$ and an initial spin period of $P\approx13.18\,\mathrm{ms}$. This result is consistent with most newborn millisecond magnetars detected in SNe~II \citep[e.g.][]{SN2017ati_Peng2026A&A...709A.156P}. However, we note that the ejecta mass inferred from this model is $M_{\rm ej}\approx0.21\,\mathrm{M_\odot}$, which is inconsistent with observational features. Because the spectroscopic features of SN~2025dys are helium-rich or even hydrogen-rich (see Sect.~\ref{sec:spec}), such a low ejecta mass is physically incompatible with an event like SN~2025dys. Therefore, we do not favour the magnetar-driven model as the explanation for the luminous light curve evolution of SN~2025dys.

\subsection{X-ray observations}\label{sec:phot:xray}

We obtained data from the \textit{Swift}-XRT instrument and processed the observations using the online tools provided by the UK Swift Science Data Centre (UKSSDC)\footnote{See \url{https://www.swift.ac.uk/user_objects/}} along with the \textsc{heasoft} software version 6.35.2 \citep{2014ascl.soft08004N}. We performed source detection within the field of SN~2025dys before generating light curves and estimating fluxes, aiming to confirm the statistical presence of the transient. To verify the origin of the detected emission, we executed two separate detection routines: one covering the epoch between approximately $2$ and $4$~days after the explosion (corresponding to the first four observations) and another at $\gtrsim20$~days after the explosion (corresponding to the last four observations). The first four observations resulted in a detection at coordinates $\mathrm{RA}=10^{\mathrm{h}}50^{\mathrm{m}}19^{\mathrm{s}}.96$, $\mathrm{Dec}=-22\arcdeg19\arcmin44\farcs9$. This location is offset by $4\farcs25$ from the optical position of SN~2025dys; this offset falls well within the positional uncertainty of $5\farcs4$. This spatial coincidence strongly suggests that the detected X-ray source corresponds to the early-phase emission of SN~2025dys. In contrast, no significant X-ray emission was detected in the field during the epoch $\gtrsim20$~days after the explosion. The early-time detection yielded a total of $20$ counts against an expected background of approximately $1.86$ counts. This result corresponds to a signal-to-noise ratio (SNR) of approximately $2.93$.

By binning the first four observations, we measured a net count rate of $(3.943\pm1.100)\times10^{-3}\,\mathrm{counts\,s^{-1}}$, which corresponds to a SNR of approximately $3.7$. For the subsequent calculations, we adopted a Galactic neutral hydrogen column density of $N_{\rm H}=5.15\times10^{20}\,\mathrm{cm^{-2}}$ and a luminosity distance of $58.1\,\mathrm{Mpc}$, as discussed in Sect.~\ref{sec:disc}. Assuming a power-law spectrum with a photon index of $\Gamma=2$, following the method utilized by \cite{SN2025aico_Zhao2026arXiv260710671Z}, yields an unabsorbed $0.3$--$10.0$~keV X-ray luminosity of $L_{\rm X} \approx 6.38 \times 10^{40}\,\mathrm{erg\,s^{-1}}$. Alternatively, adopting a thermal bremsstrahlung spectrum with a temperature of $kT = 10$~keV, as implemented by \cite{SN2024iss_Chen2026AA}, results in a luminosity of $L_{\rm X} \approx 7.55 \times 10^{40}\,\mathrm{erg\,s^{-1}}$. We noted that the majority of the detected photons possess energies $\lesssim 1$~keV, indicating that the early X-ray emission of SN~2025dys is predominantly soft.

Crucially, this early X-ray detection poses a stringent constraint on the geometry of the circumstellar medium. If we assume a spherically symmetric geometry for the massive circumstellar medium derived from our light curve modeling, the equivalent hydrogen column density would be extremely high ($N_{\rm H} \approx 10^{26}\,\mathrm{cm^{-2}}$). Under such dense conditions, it is less likely for soft X-ray photons to escape. A natural explanation to resolve this issue is an intrinsically asymmetric distribution of the dense material. Consistent with our previous fitting results, the dense material located at $R \gtrsim 10^{14}\,\mathrm{cm}$ is highly likely to be distributed in an equatorial ring or in a clumpy configuration. Within the inner cavity, the environment likely consists of a steady stellar wind characterized by a significantly lower density. In this geometrical scenario, the detected X-ray emission is generated by the shock interacting with this inner wind region, and the photons escape along the relatively unobscured polar directions. To quantify the properties of this inner wind, we applied the model of \cite{Fransson1996ApJ...461..993F}. Adopting a typical steady-wind velocity of $10\,\mathrm{km\,s^{-1}}$, a shock velocity of $5200\,\mathrm{km\,s^{-1}}$ derived from our spectral analysis, and an average observation epoch of $\sim 3$~days after explosion, we deduce a mass-loss rate of approximately $5\times10^{-4}\,\mathrm{M_\odot\,yr^{-1}}$. This value represents the mass-loss rate of the inner region close to the progenitor, rather than the outer dense shell. This derived rate is remarkably consistent with typical stellar winds originating from red or yellow supergiant progenitors \citep{Smith2014ARAA}, thereby providing independent support for the asymmetric circumstellar geometry we propose.

\section{Spectroscopy}
\label{sec:spec}
\subsection{Spectroscopic evolution}
\begin{figure*}[htbp]
\centering
\begin{minipage}[c]{0.52\textwidth}
  \centering
  \includegraphics[width=\linewidth]{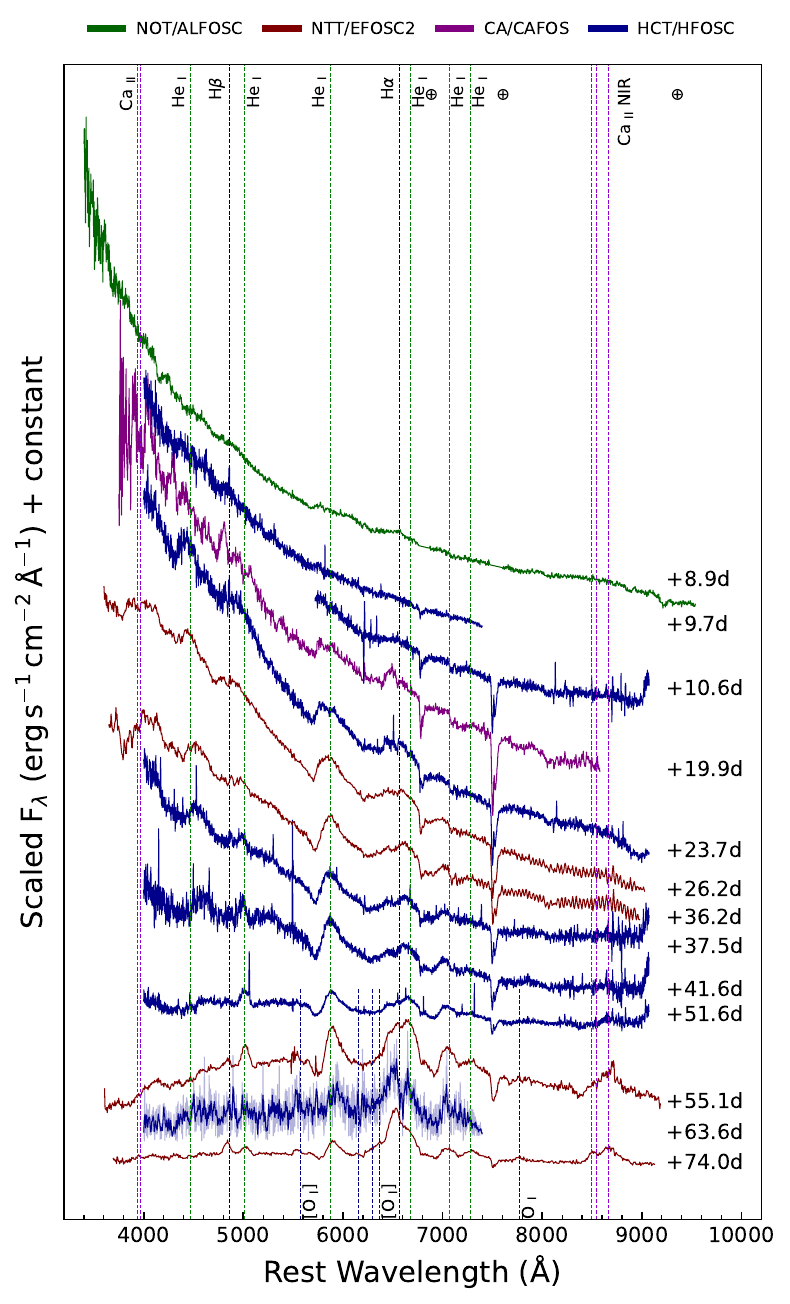}
\end{minipage}\hfill
\begin{minipage}[c]{0.47\textwidth}
  \centering
  \includegraphics[width=\linewidth]{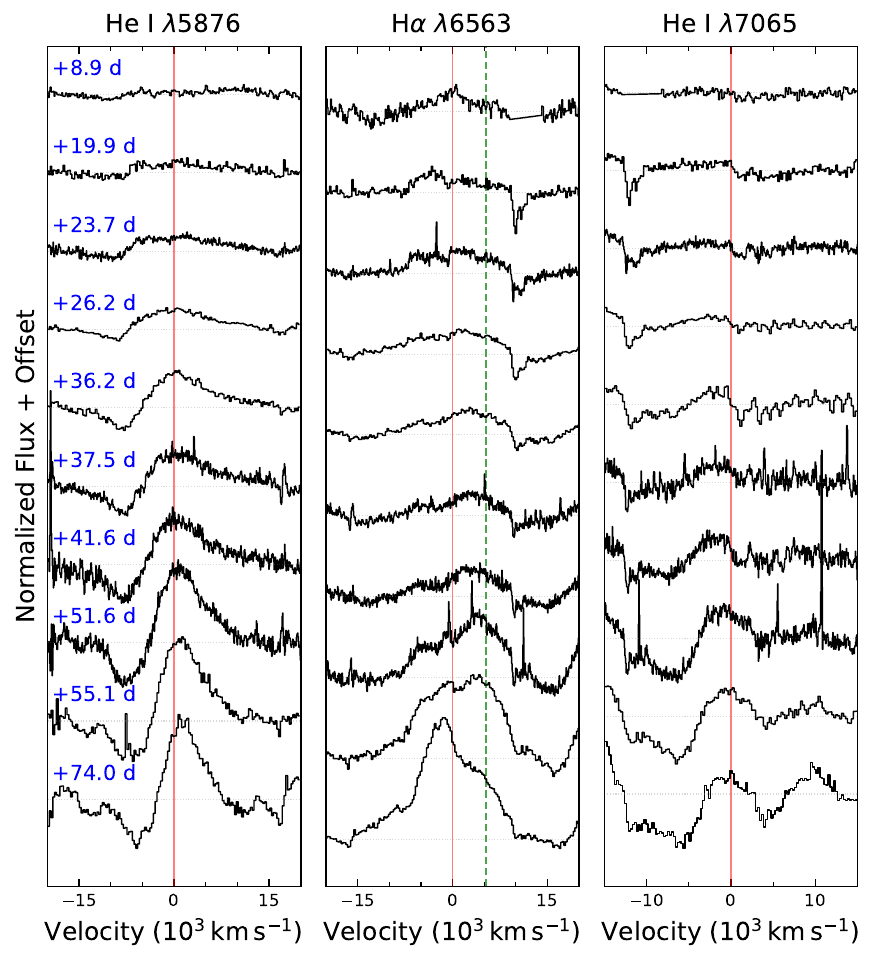}
  \includegraphics[width=\linewidth]{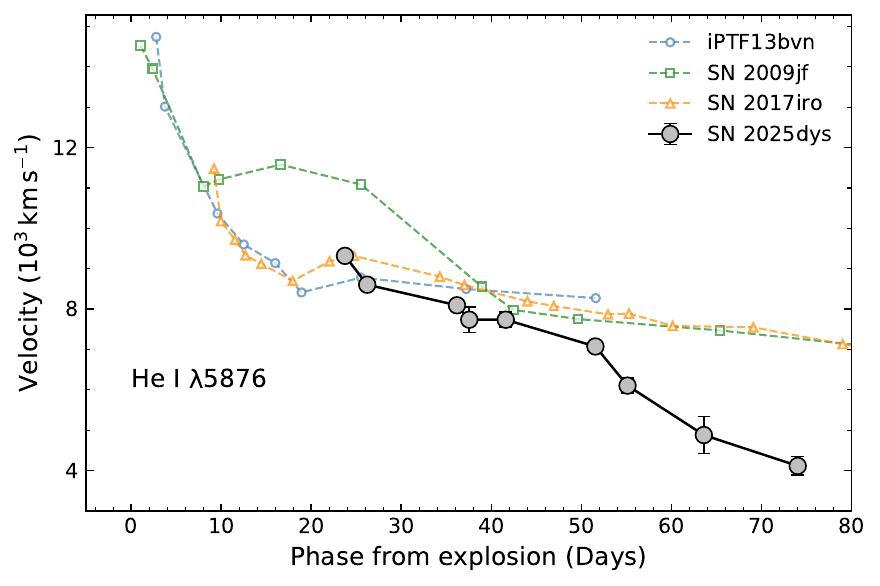}
\end{minipage}\hfill
\caption{Spectroscopic sequence of SN~2025dys (left panel). The spectra are scaled and shifted by the appropriate offset to maintain clarity. Low-S/N spectra are re-binned for clarity, with the original spectra shown in lighter colours. The right-side marked epochs are relative to the estimated explosion epoch. Spectra from different facilities are marked with different colours. The evolution of H$\alpha$ and \ion{He}{i} lines is shown on the upper right, and the \ion{He}{i}~$\lambda5876$ line velocity evolution of different events is displayed in the bottom right panel.}
\label{fig:spec:evo}
\end{figure*}
Fig.~\ref{fig:spec:evo} presents the low- and medium-resolution spectroscopic sequence of SN~2025dys, and Section~\ref{sec:app:ctf:red} details the spectroscopic data reduction. The data cover the period from $+8.9$~days to $+74.0$~days after the explosion, encompassing the rise, peak luminosity, decline, and radioactive tail phases. During the early phases ($+8.9$\,--\,$+10.6$~days), the spectra exhibit a strong, nearly featureless blue continuum, with only a few possible broad electron-scattering profiles superimposed, similar to those of SN~1979C \citep{Branch1981ApJ...244..780B}. The corresponding blackbody temperature is approximately $13,230^{+1,400}_{-1,300}$~K, which is consistent with events experiencing strong early circumstellar interaction \citep[see, e.g.,][]{Charalampopoulos2025A&A...700A.138C}\footnote{Note that this temperature is lower than that estimated in Sect.~\ref{sec:phot:lc}. This discrepancy is due to the lack of UV and NUV coverage in the spectroscopic data. Thus, the temperature derived here is likely underestimated, and we adopt the result presented in Sect.~\ref{sec:phot:lc} in the following analysis.}. However, these early spectra lack the prominent narrow emission lines that indicate strong interaction, such as those observed in other early-interaction events including SN~1998S and SN~2017ahn \citep{SN1998S_Anupama2001A&A...367..506A, Tartaglia2021ApJ...907...52T}.

At $+19.9$~days, discrete spectral features begin to emerge. The spectra display prominent, blueshifted Balmer emission lines, such as H$\alpha$ and H$\beta$. We measured the FWHM of the H$\beta$ emission line by fitting a Gaussian profile to the emission line, resulting in a velocity of $\sim3000\,\mathrm{km\,s^{-1}}$, which is consistent with the typical CDS velocity. From $+23.7$~days to $+41.6$~days, the continuum becomes redder, indicating a drop in the blackbody temperature from $11,770\pm1,100$~K to $8,050\pm100$~K. The H$\alpha$ feature exhibits a broad, flat-topped profile lacking a prominent P~Cygni absorption component. This morphology likely originates from emission within a shocked cold dense shell (CDS). Furthermore, the spectrum presents prominent P~Cygni profiles for the \ion{He}{i}~$\lambda5876$ and \ion{He}{i}~$\lambda7065$ lines. In addition, other helium lines, such as \ion{He}{i}~$\lambda5016$ and \ion{He}{i}~$\lambda6678$, show emission profiles. These features suggest that the ejecta are helium-rich, similar to the ejecta of SNe~IIb \citep[see, e.g.][]{SN2024iss_Chen2026AA, SN2022ngb_Zhao2026A&A}. The presence of broad \ion{He}{i} P~Cygni profiles marks the signature of fast-expanding ejecta. This behaviour resembles the photospheric phase of SN~2018gjx, although the CSM of SN~2018gjx is helium-rich \citep{SN2018gjx_Prentice2020MNRAS.499.1450P}. After $+51.6$~days, the evolution enters a transitional phase during which the photosphere begins to recede, as shown previously in Fig.~\ref{fig:phot:bolcolor}. The continuum becomes redder, and the blackbody temperature remains at $\sim 6,020\pm100$~K. The P~Cygni absorption components become less prominent, while \ion{He}{i} and \ion{Ca}{ii} near-infrared triplet emission lines emerge. Similar features appear during the transitional or nebular phases of SNe~IIb \citep{Jerkstrand2015aap}. Furthermore, Balmer emission lines, including H$\alpha$ and H$\beta$, increase in prominence. These features eventually dominate the spectra, reaching intensities comparable to the \ion{He}{i} lines. This enhancement could be a signature of late-phase interaction, with emission originating from the outer CDS.

After $+63.6$~days, the H$\alpha$ and \ion{He}{i} emission lines can be resolved separately, and the H$\alpha$ profile exhibits a blueshifted peak. Absorption features nearly disappear from the spectrum, and the [\ion{O}{i}]~$\lambda5577$ emission line emerges with a blueshifted peak. At $+74.0$~days, emission from the [\ion{O}{i}]~$\lambda\lambda6300,\,6364$ doublet appears on the blue wing of the H$\alpha$ profile, marking the onset of the nebular phase. Simultaneously, the strength of the [\ion{O}{i}]~$\lambda5577$ line fades, indicating a cooling of the emitting region \citep{Elmhamdi2004A&A}. The H$\alpha$ emission becomes asymmetric, and an apparent red wing emerges. This feature blends with the \ion{He}{i}~$\lambda6678$ line, which is expected to have an intensity similar to other \ion{He}{i} transitions. The peak of H$\alpha$ presents a clear blueshift, consistent with other Balmer lines such as H$\beta$. Finally, the [\ion{Ca}{ii}]~$\lambda\lambda7291,\,7324$ doublet emerges in the red portion of the spectrum, potentially blending with the \ion{He}{i}~$\lambda7281$ line. However, because we do not notice any prominent \ion{He}{i}~$\lambda7281$ emission in the earlier spectra, we assume that this feature is primarily produced by calcium.

Furthermore, we present the velocity evolution derived from the P~Cygni profile of \ion{He}{i}~$\lambda5876$. The velocity was determined by identifying the minimum of the absorption trough. For low S/N spectra, the minimum was measured directly, whereas for higher S/N data, a Gaussian profile was fitted to the absorption feature. The P~Cygni profile emerged at $+23.7$~days, showing an expansion velocity of approximately $9,300\,\mathrm{km\,s^{-1}}$. Between $+20$ and $+40$~days, the velocity decreased gradually, reaching $7,800\pm120\,\mathrm{km\,s^{-1}}$ at $+41.6$~days. This behaviour closely tracks the typical evolution observed in a comparison sample of SNe~Ib. However, a significant divergence appeared at $t\gtrsim50$~days, as the decline rate of the \ion{He}{i}~$\lambda5876$ velocity increased. The velocity dropped rapidly from $7,100\pm120\,\mathrm{km\,s^{-1}}$ to $4,100\pm230\,\mathrm{km\,s^{-1}}$, whereas the comparison objects exhibited a much more stable velocity evolution during this phase. This sudden drop can be naturally explained by the rapid recession of the photosphere into the inner layers of the ejecta. Because the outer ejecta had previously been swept up and compressed by a reverse shock arising from the early interaction with a H-rich CSM, the remaining inner core intrinsically possessed a much lower homologous velocity. Furthermore, the evolution of the blackbody radius (presented in Sect.~\ref{sec:phot:lc}) indicates that the photospheric radius reached a maximum at approximately $+50$~days before beginning to shrink. Concurrently, the temperature reached a floor of approximately $5,700$~K, which corresponds to the recombination temperature of H-rich materials \citep[see,][]{Nagy2014AA}, and the velocity is also consistent with the ejecta velocity which we derived from light curve modelling (see Sect.~\ref{sec:phot:lcmdl}, approximately $5200\,\mathrm{km\,s^{-1}}$). These photometric trends are consistent with the hypothesis proposed above, indicating that the photosphere had begun to recede deeply into the inner ejecta.

\subsection{Spectral modelling and comparisons}
\label{sec:spec:mod}
\subsubsection{Electron scattering profile in early spectrum}
\label{sec:spec:mod:ec}
At $+8.9$~days, the spectrum is dominated by a featureless blue continuum and a marginal H$\alpha$ emission line exhibiting broad electron-scattering wings. \citet{SN2020able_Tartaglia2026arXiv260807661T} suggest that the profile of electron-scattering wings can be used to constrain the mass-loss rate of the progenitor and estimate the mass of the unshocked CSM. Here, we utilize a simple Monte Carlo simulation, which is implemented via a modified version of the \textsc{escatter} code \citep{Brennan2024A&A...684L..18B, Brennan2024ascl.soft01020B}. We refine the handling of geometric boundaries and incorporate the effect of wind velocity\footnote{However, because the wind velocity in our case is limited to a range of $\sim10$--$100\,\mathrm{km\,s^{-1}}$, it has a negligible effect on the profile of the electron-scattering wings}. We normalize the spectrum of SN~2025dys at $+8.9$~days and subtract the continuum baseline. Subsequently, we scale the flux level to a range between $0$ and $1$, and isolate the $\pm12,000\,\mathrm{km\,s^{-1}}$ velocity region around the H$\alpha$ emission line.

We adopt a uniform electron temperature for the environment, which is approximated by the blackbody temperature of $17,500$~K. In the Monte Carlo simulation, we assume that photons are emitted from the continuum-emitting boundary\footnote{Here, we use $R_{\rm bb}$ at $+8.9$~days as a proxy for the inner boundary.} at $+8.9$~days with zero initial Doppler shift. The outer boundary of the CSM is derived from the first breakpoint in the evolution of the blackbody radius, which represents the shock sweeping through the entire CSM. We apply a piecewise linear fit with three segments to the blackbody radius of SN~2025dys, as discussed in Sect.~\ref{sec:phot:lc}, yielding a first breakpoint at $18.5\pm1.1$~days with a corresponding radius of $1.05\times10^{15}\,\mathrm{cm}$. As a consistency check, we note that a broad P~Cygni profile emerges in the spectrum at $+19.9$~days, indicating the signature of the expanding ejecta. This observation is consistent with our deduction from the blackbody radius fit that the CSM is fully shocked by this epoch \citep[see][]{SN2020able_Tartaglia2026arXiv260807661T}. Thus, the simulation region spans the radial range of $(R_{\rm in},\,R_{\rm out}) = (5.5,\,10.5)\times10^{14}\,\mathrm{cm}$. We construct a simulation grid covering a CSM optical depth range from $\tau_{\rm CSM}=2$ to $20$ in steps of $0.1$. To reproduce both the broad wings and the central peak, the final model combines the simulated electron-scattering profile with a Gaussian component, which accounts for the unscattered (or weakly scattered) line core emitted from the outer CSM layers. We employ a $\chi^{2}$ minimization technique to identify the best-matching model. Considering the hydrogen-rich environment of the CSM, we adopt a typical optical opacity for SNe~II \citep{Anderson2014ApJ...786...67A} of $\kappa_{\rm opt}=0.34\,\mathrm{cm^{2}\,g^{-1}}$, which is consistent with our previous analysis.

\begin{figure}[htbp]
\centering
\includegraphics[width=\linewidth]{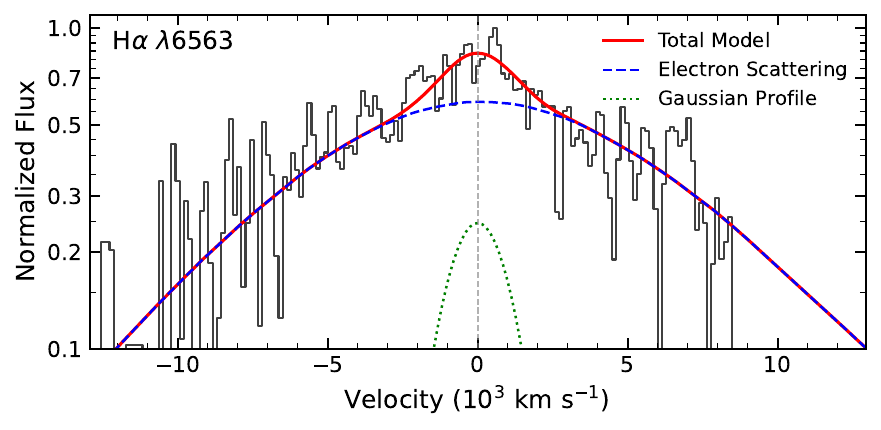}
\caption{Electron scattering model of SN~2025dys. The red continuous line and blue dashed line represent the total model and the electron scattering component, respectively. The unscattered Gaussian profile is shown with a green dotted line (see Sect.~\ref{sec:spec:mod:ec}).}
\label{fig:spec:esmdl}
\end{figure}

The fitting results are presented in Fig.~\ref{fig:spec:esmdl}. The best-fit electron scattering model yields an optical depth of $\tau_{\mathrm{CSM}}=13$. Considering that the optical depth represents the CSM ahead of the continuum-emitting boundary ($R_{\rm in}$), this result is reasonable at the epoch of $+8.9$~days, as the boundary has expanded into the surrounding environment. The corresponding unshocked CSM ahead of $R_{\rm in}$ has a mass of $M=0.14\,\mathrm{M_\odot}$. Assuming a steady-wind density profile ($s=2$) and a typical eruptive mass-loss wind velocity of $v_{\mathrm{wind}}=100\,\mathrm{km\,s^{-1}}$, this yields a mass-loss rate of $\dot{M}=0.09\,\mathrm{M_\odot\,yr^{-1}}$. This result is consistent with eruptive mass loss from a luminous blue variable (LBV) \citep{Smith2017hsn..book..403S}. However, because the light-curve fitting results yield a less massive progenitor, an LBV origin is unlikely. Therefore, the mass-loss history of SN~2025dys is more likely driven by a nuclear flash (e.g., a silicon flash; see \citealt{Chiba2025MNRAS.542.2353C, Woosley2019ApJ...878...49W}) or caused by Roche-lobe overflow. In addition, the superwind of an RSG is also a possibility that cannot be completely ruled out \citep[see,][]{Cai2026arXiv260630390C}. Taking these scenarios into account, where the wind velocity is typically around $50\,\mathrm{km\,s^{-1}}$, yields a mass-loss rate of $\dot{M}=0.05\,\mathrm{M_\odot\,yr^{-1}}$. Ultimately, considering all possible scenarios for SN~2025dys, we derive a lower limit for the mass-loss rate of $\dot{M} \gtrsim 0.01\,\mathrm{M_\odot\,yr^{-1}}$.

\subsubsection{Spectral comparisons of other SNe~II and synthetic spectra}
\label{sec:spec:mod:cmp}

\begin{figure*}[htbp]
\centering
\begin{minipage}[c]{0.51\textwidth}
  \centering
  \includegraphics[width=\linewidth]{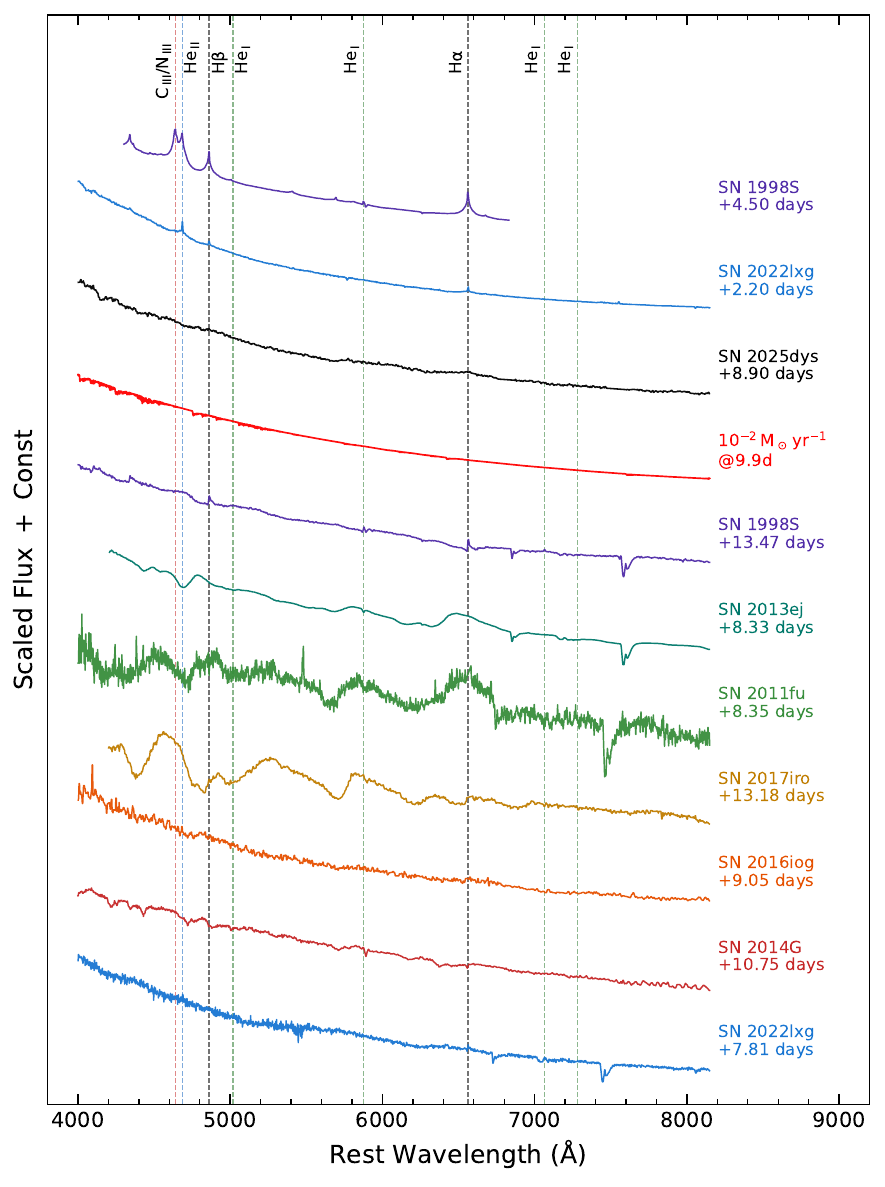}
\end{minipage}\hfill
\begin{minipage}[c]{0.48\textwidth}
  \centering
  \includegraphics[width=\linewidth]{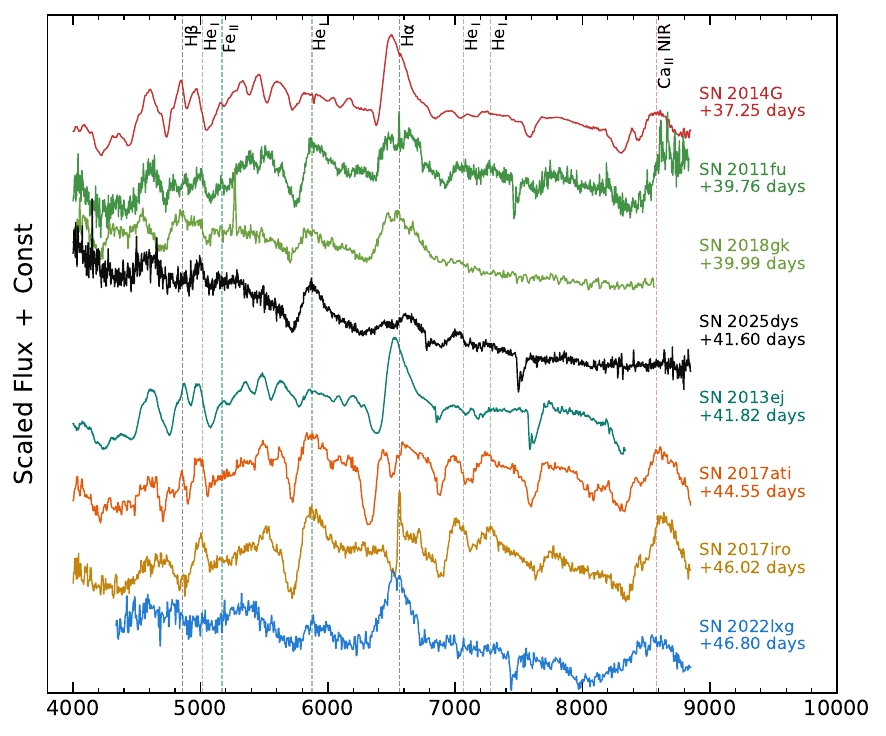}
  \includegraphics[width=\linewidth]{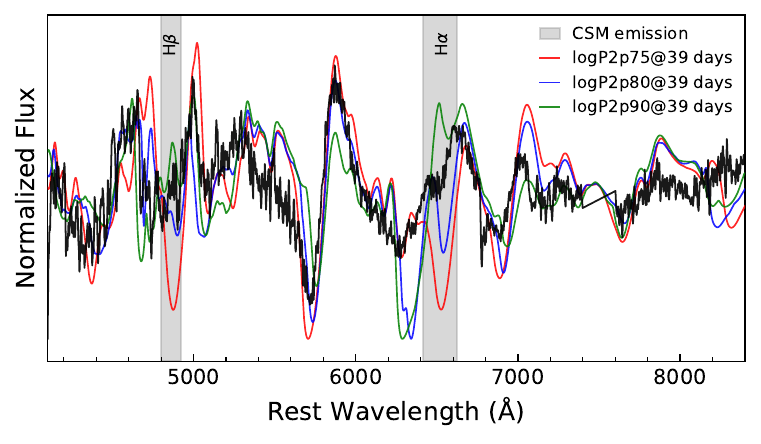}
\end{minipage}\hfill
\caption{Spectroscopic comparisons of SN~2025dys in the early and post-peak phases. The comparisons are marked with different colours with an offset. Important spectral lines are also marked, and the corresponding labels indicate the respective phases. A model spectrum from \citet{Dessart2017A&A...605A..83D} is also displayed in the left panel (for details, see Sect~\ref{sec:spec:mod:cmp}). The synthetic spectra are shown with different colours, and the region for the possible CSM emission is denoted in the grey region. Phases are calculated from the estimated explosion epoch.}
\label{fig:spec:cmpem}
\end{figure*}

We compared the spectra of SN~2025dys to those of similar events, including interacting Type~II supernovae and hydrogen-rich events powered by magnetars, as well as to synthetic spectral models. These are presented in Fig.~\ref{fig:spec:cmpem}. In the early phase at $\sim$+8.9~days, SN~2025dys did not exhibit prominent narrow emission lines, but rather displayed a nearly featureless continuum with only marginally broad electron-scattering wings. Overall, the early spectrum is similar to spectra of events at comparable phases, such as SN~2014G and SN~2016iog, and also resembles models with a mass-loss rate of approximately $10^{-2}\,\mathrm{M_\odot\,yr^{-1}}$ \citep{Dessart2017A&A...605A..83D}. The featureless nature of the spectra could result from a dense and optically thick CSM, which causes photons to scatter rather than escape to form emission lines \citep[see, e.g.,][]{Peng2026A&A...705A.104P}. 
An alternative explanation is that narrow lines might have existed in the very early phase, but we are missing these very early detections.
Such evolution is observed in the early spectra of SN~1998S and SN~2022lxg, both of which exhibit narrow lines at $t \lesssim 3$~days, an epoch much earlier than our first spectroscopic detection. Furthermore, in Sect.~\ref{sec:phot:lc}, we noted that the intrinsic colour $(g-r)_0$ is relatively red in the early phase, which could alternatively be explained by strong early H$\alpha$ emission. Therefore, the possibility of having missed early narrow-line emission cannot be completely ruled out. In comparison to events with weak or no prominent interaction, such as SN~2011fu, SN~2017iro, and SN~2013ej, broad P~Cygni profiles are typically already visible at this phase, whereas they remain absent in the spectrum of SN~2025dys. This suggests that the optical depth of the continuum remains high in SN~2025dys.
\begin{figure}[htbp]
\centering
\includegraphics[width=\linewidth]{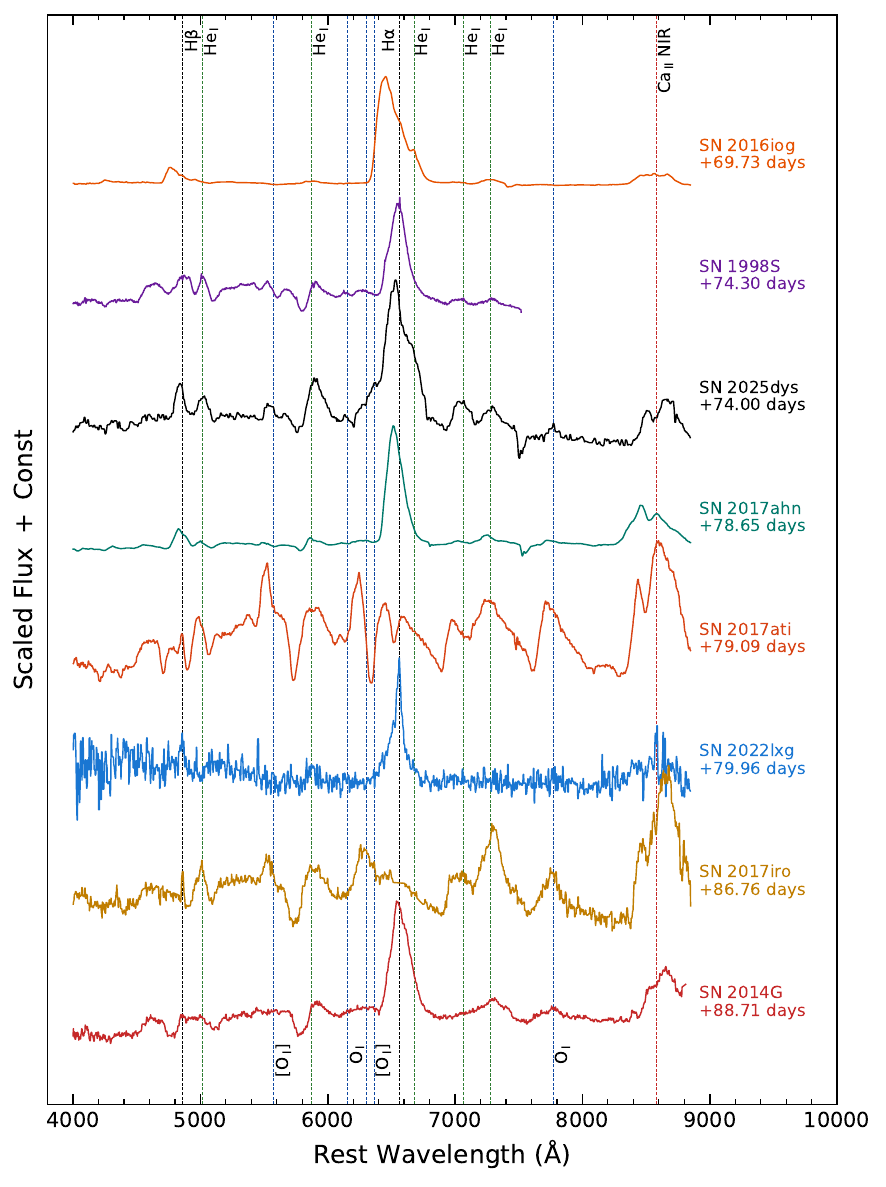}
\caption{Spectral comparisons of SN~2025dys in the late phase with other similar or relevant events. Prominent lines are marked with corresponding markers. Different targets are present in different colours. Phases are calculated from the estimated explosion epoch.}
\label{fig:spec:cmplate}
\end{figure}

The post-peak spectrum of SN~2025dys at $+41.6$~days after explosion still shows a relatively bluer continuum compared to all other reference events. This indicates that the ejecta of SN~2025dys remain at a high temperature, suggesting sustained energy input. A similar feature was also observed in SN~2018gk, a Type~IIb event powered by a central engine \citep{Wang2023ApJ...948..138W}. However, considering the intermediate-width Balmer emission lines discussed in Sect.~\ref{sec:spec}, along with the flat H$\alpha$ profile lacking a prominent P~Cygni signature, we do not favor the central-engine scenario. A central engine would typically produce broad P~Cygni profiles for Balmer lines, as seen in SN~2018gk and SN~2017ati \citep{SN2017ati_Peng2026A&A...709A.156P}, rather than these intermediate-width features. In this phase, the spectral features of SN~2025dys resemble those of hydrogen-poor targets, such as SN~2017iro, where the spectra are dominated by \ion{He}{i} P~Cygni profiles. In fact, SN~2025dys appears to be even more deficient in hydrogen than typical Type~IIb events like SN~2011fu and SN~2018gk \citep{Kumar2013mnras}. This behaviour is in stark contrast to hydrogen-rich interacting events such as SN~2022lxg, SN~2014G, and SN~2013ej. At the redder end of the spectrum of SN~2025dys, the \ion{Ca}{ii} near-infrared (NIR) triplet is absent, a feature that is typically present in most other compared events. One possible explanation is that, due to the high temperature in SN~2025dys, most of the calcium is ionized to \ion{Ca}{iii} rather than remaining as \ion{Ca}{ii}. This is consistent with the high-temperature scenario proposed for SN~2018gk.

Furthermore, we compared the spectrum of SN~2025dys to the models of \cite{Dessart2016MNRAS, Dessart2024A&A...685A.169D}. We found that SN~2025dys closely resembles the \texttt{logP2p75} model, which represents a Type~Ib supernova. In contrast, models with a higher hydrogen content, such as \texttt{logP2p80} and \texttt{logP2p90}, significantly overpredict the hydrogen features compared to our observations. Specifically, both the \texttt{logP2p80} and \texttt{logP2p90} models display a prominent H$\alpha$ P~Cygni profile, characterized by a deep, blueshifted absorption trough. However, SN~2025dys exhibits no such prominent H$\alpha$ absorption. Instead, the observed spectrum aligns well with the \texttt{logP2p75} model in the $6000$--$6400\,\mathrm{\AA}$ region. Additionally, the \texttt{logP2p90} model underestimates the intensity of the \ion{He}{i}~$\lambda5876$ line. This further suggests that the underlying ejecta of SN~2025dys are highly deficient in hydrogen, akin to Type~IIb models with minimal hydrogen or even Type~Ib supernovae. The only notable discrepancy with the \texttt{logP2p75} model arises near the rest wavelengths of H$\alpha$ and H$\beta$. In the \texttt{logP2p75} model, these regions correspond to the absorption components of the \ion{He}{i}~$\lambda6678$ and \ion{He}{i}~$\lambda5016$ P~Cygni profiles, respectively. Given the context of circumstellar interaction, this inconsistency can be naturally explained by H$\alpha$ and H$\beta$ emission originating from the CDS. This emission is centered near the rest wavelengths and possesses an intermediate width of $\sim3000\,\mathrm{km\,s^{-1}}$. This intermediate-width emission effectively fills in the underlying \ion{He}{i} absorption troughs predicted by the model, resulting in a flatter observed profile. Crucially, if the ejecta themselves possessed significant hydrogen, the resulting high-velocity H$\alpha$ absorption\footnote{$v_{\rm H}\gtrsim 10000\,\mathrm{km\,s^{-1}}$ typical for stripped-envelope (especially SNe~IIb) supernovae} would be shifted far to the blue. Such a broadly blueshifted trough could not be filled by the relatively narrow, rest-wavelength emission from the CDS \citep[see, e.g.,][]{SN2022ngb_Zhao2026A&A, SN2024aecx_Zou2026ApJ...997...77Z}. Therefore, based on these synthetic spectral comparisons, we conclude that the inner ejecta of SN~2025dys must be intrinsically hydrogen-poor, mirroring Type~IIb or Type~Ib supernovae. This dual nature firmly establishes SN~2025dys as a transitional event, which may share a similar explosion scenario to the events like SN~2014C, SN~2019oys and SN~2019tsf, but having a stronger signature of the interaction in the early time rather in the late phase \citep{SN2014C_2025ApJ...978..163Z, SN2014C_Milisavljevic2015ApJ...815..120M, Sollerman2020A&A...643A..79S}.

In the late-phase spectra of SN~2025dys ($t \gtrsim 50$~days), Balmer emission lines re-emerged, causing the overall spectral morphology to more closely resemble those of hydrogen-rich events. This indicates a resurgence of circumstellar interaction. Such behaviour closely follows the evolutionary path of events like SN~2018gjx, which exhibited interaction signatures at early phases, subsequently transitioned to an ejecta-dominated phase, and eventually displayed renewed interaction signatures at late times \citep{SN2018gjx_Prentice2020MNRAS.499.1450P}. Accordingly, we compared our final spectrum with spectra of reference events at epochs similar to that of SN~2025dys. The results are presented in Fig.~\ref{fig:spec:cmplate}. During this transitional or early nebular phase, SN~2025dys shares similarities with hydrogen-rich interacting events such as SN~2014G, SN~1998S, and SN~2022lxg. However, SN~2025dys exhibits stronger \ion{He}{i} emission and P~Cygni profiles, which are characteristic of transitional Type~IIb or Type~Ib events like SN~2017ati and SN~2017iro \citep{SN2017ati_Peng2026A&A...709A.156P, SN2017iro_Kumar2022ApJ...927...61K}. Furthermore, the \ion{Ca}{ii} near-infrared triplet began to emerge. This suggests that as the ejecta cooled and the flux of high-energy ionizing photons from the interaction region decreased, calcium recombined from \ion{Ca}{iii} back to \ion{Ca}{ii}. Additionally, several oxygen lines appeared in the spectrum of SN~2025dys, including forbidden transitions such as [\ion{O}{i}]~$\lambda\lambda6300, 6364$. The emergence of these lines in such an early phase indicates that the progenitor of SN~2025dys was likely resembled the highly stripped progenitors of events like SN~2017iro. Finally, the emission peak of [\ion{O}{i}]~$\lambda5577$ exhibits a systematic blueshift of approximately $50\,\mathrm{\AA}$, which might suggest an asymmetric explosion. However, caution is warranted, as residual opacity obscuring the receding side of the ejecta could also produce a blueshifted profile, and potential blending with nearby \ion{Fe}{ii} lines further complicates this interpretation.

Furthermore, the observed H$\alpha$ profile of SN~2025dys is highly complex. By comparing it with reference targets, we identified that the bump on the blue edge of the H$\alpha$ emission is contributed by the [\ion{O}{i}]~$\lambda\lambda6300, 6364$ doublet. The most prominent peak is driven by H$\alpha$ emission originating from the expanding shocked cold dense shell. Meanwhile, the red wing is likely a blended contribution of both \ion{He}{i}~$\lambda6678$ and H$\alpha$ emission. Cross-checking with the intensities of other \ion{He}{i} lines reveals that helium emission alone is insufficient to account for the entire flux of this red bump. Consequently, SN~2025dys presents an asymmetric H$\alpha$ profile featuring a blueshifted peak, a characteristic also observed in SN~2016iog and SN~2017ahn. One explanation for this morphology is an intrinsic asymmetry in the CSM, which is most likely distributed in a ring-like or clumpy geometry that suppresses the receding (redshifted) emission component. Another possible explanation is the formation of dust, a mechanism frequently utilized to explain similar profiles in SN~2016iog and Type~IIn events such as SN~2019vxm \citep{Peng2026A&A...705A.104P, Cai2026arXiv260630390C, Rameshan_2026R}. However, at an epoch of $\sim75$~days, it is relatively early for substantial dust formation. Moreover, no evidence supporting early dust formation has been found in similar events like SN~2017ahn. More crucially, the formation of dust would typically obscure the receding emission across all spectral lines; yet, the \ion{He}{i} profiles in our observations remain nearly symmetric. Therefore, we favor the explanation of an asymmetric CSM, which is also consistent with our previous deductions regarding the asymmetric structure of the progenitor. Nevertheless, we must exercise caution; without near-infrared observations, the possibility of dust formation cannot be completely ruled out.

\section{Discussion and conclusion}
\label{sec:discon}
In this article, we present the photometric and spectroscopic observations and analysis of the Type~II supernova SN~2025dys. The evolution of the light curve suggests SN~2025dys to be a high-luminosity event, with a peak luminosity comparable to those of SN~1998S and SN~2022lxg. The photometric evolution of SN~2025dys exhibits a rise time of $\sim18$~days, which is longer than that of typical early-interacting Type~II events, yet it presents a rapid evolutionary decline following the peak luminosity. After $\sim70$~days, the light curve of SN~2025dys exhibits a linear decline, which corresponds to the RD phase. The evolution of the spectral energy distribution and colour suggests a prominent photon diffusion process during the early rising phase, characterized by initial heating and a relatively red intrinsic colour. 

The light curve fitting results obtained using \textsc{mosfit} suggest that the photometric evolution of SN~2025dys can be naturally explained by a combination of radioactive decay and CSI \citep[see, e.g.,][]{Guillochon2018ApJS..236....6G, Rameshan_2026R}. The best-fit parameters yield a massive, optically thick CSM with a mass of $M_{\rm CSM}\approx0.5$--$0.75\,\mathrm{M_\odot}$ and a relatively high density of $10^{-11}$--$10^{-10}\,\mathrm{g\,cm^{-3}}$\footnote{Since $s=2$ shows a more reasonable result, thus we report the posterior distribution of $s=2$. For $s=0$, the density is $\sim10^{-12}\,\mathrm{g\,cm^{-3}}$.}. Regarding the ejecta, the fitting results indicate a partially stripped progenitor that produced a low-mass ejecta of $M_{\rm ej}\approx2.5\,\mathrm{M_\odot}$, a value significantly lower than those of typical hydrogen-rich Type~II supernovae. The kinetic energy is also suggested to be relatively low, at approximately $0.5$~foe. Finally, the synthesized nickel mass is estimated to be $M_{\rm Ni}\approx0.045\,\mathrm{M_\odot}$, which is consistent with the typical range observed in Type~II SNe, as well as in some low-luminosity events.

The spectroscopic observations of SN~2025dys provide further insights into the detailed properties of the progenitor. In the early phase, SN~2025dys exhibited a nearly featureless and extremely blue continuum, displaying only marginal electron-scattering wings around H$\alpha$. The best-fit model of the line profile indicates a high mass-loss rate of $\dot{M} \gtrsim 0.01\,\mathrm{M_\odot\,yr^{-1}}$ and an optical depth of $\tau \approx 13$ for the unshocked material. This is consistent with the presence of a dense and massive CSM, which naturally drives the photometric evolution described in our light curve modeling. Distinct spectral features began to emerge at $+19.9$~days, marked by the appearance of intermediate-width Balmer emission lines alongside broad P~Cygni profiles of \ion{He}{i}. Physically, the intermediate-width features likely originate from the shocked cold dense shell, while the broad lines trace the underlying ejecta. The continuum temperature of SN~2025dys remained high around $\sim15,000\,\mathrm{K}$, eventually declining to $\sim 5700$~K at $\sim 60$~days. This temperature is significantly higher than those of early-interacting Type~II events\footnote{Typically around $6,000$--$12,000$~K at comparable post-peak epochs.} and typical stripped-envelope supernovae, indicating a sustained additional energy input. 

Furthermore, the spectral features observed between $\sim 20$ and $50$~days reveal the properties of the inner ejecta. These observations suggest that the inner ejecta are highly hydrogen-poor, closely resembling those of Type~IIb or even Type~Ib supernovae, a conclusion that is entirely consistent with the results of our light curve modeling. At later epochs, the Balmer lines strengthened, indicating ongoing circumstellar interaction, and exhibited a systematic blueshift in their emission peaks. Additionally, a prominent bump appeared on the red wing of the H$\alpha$ line, resulting in a broad and highly asymmetric profile. This bump could partially arise from a blended contribution of H$\alpha$ and \ion{He}{i}~$\lambda6678$. Alternatively, the intrinsic asymmetry and the systematic blueshift of the Balmer profiles could be driven by a highly asymmetric distribution of the CSM or by the early formation of dust. Finally, the emergence of the [\ion{O}{i}]~$\lambda\lambda6300, 6364$ doublet suggests that the ejecta were gradually becoming optically thin, marking the transition of the supernova toward the nebular phase.

By synthesizing the photometric and spectroscopic evolution, a coherent picture of the progenitor and explosion scenario can be established. The progenitor of SN~2025dys is suggested to be a partially stripped helium star, similar to the progenitors of Type~IIb or even Type~Ib supernovae. However, unlike typical stripped-envelope events, the immediate environment of SN~2025dys is not hydrogen-poor. Instead, the circumstellar environment features an optically thick, dense, and hydrogen-rich CSM. As discussed in Sect.~\ref{sec:spec:mod:ec}, this material could have been ejected by late-stage nuclear flashes several years prior to the explosion, or driven by instabilities during RLOF in a binary system \citep{Smith2017hsn..book..403S, Woosley2019ApJ...878...49W, Chiba2025MNRAS.542.2353C}. Furthermore, based on the early detection of soft X-ray emission and the asymmetric, blueshifted Balmer profiles observed at late phases, the CSM is highly likely to be clumpy or distributed in a ring-like geometry around the progenitor. Additionally, late-time dust formation, triggered by the cooling of the shocked cold dense shell, must also be taken into account \citep[see, e.g.,][]{Peng2026A&A...705A.104P}. 

In summary, the evolution of SN~2025dys is best explained by the explosion of a low-mass, partially stripped helium star into a dense, asymmetric, hydrogen-rich CSM. This configuration leads to sustained interaction well into the late phases, accompanied by possible dust formation. Overall, the physical nature of the ejecta resembles that of Type~IIb or Ib events, with the fundamental distinction that the surrounding CSM retains a substantial amount of hydrogen.  

\section{Data availability}
UV and optical photometric measurements and corner plots of SN~2025dys are available upon reasonable request to the corresponding author. Our spectral observations are available through the Weizmann Interactive SN Data Repository (WISeREP; \citealt{Yaron2012pasp}). 

\section*{Acknowledgements}
{
\small
This work is supported by the National Key Research and Development Program of China (Grant No. 2024YFA1611603), the National Natural Science Foundation of China (NSFC, Grant Nos. 12303054, 12473047), the Yunnan Fundamental Research Projects (Grant Nos. 202401AU070063, 202501AS070078), the Yunnan Key Laboratory of Survey Science (No. 202449CE340002), the ``Science \& Technology Champion Project'' (202005AB160002) and two ``Team Projects" -- the ``Top Team'' (202305AT350002) and the ``Innovation Team'' (202105AE160021), all funded by the ``Yunnan Revitalization Talent Support Program". 
BK acknowledges support from the ``Special Project for High-End Foreign Experts", Xingdian Funding from Yunnan Province, China and the Nanjing University research grant. 
A.S. acknowledges support from the Knut and Alice Wallenberg Foundation through project “Gravity Meets Light”.
JWZ acknowledges support from the Graduate Research Innovation Fund of Yunnan University. 
YPY acknowledges support from the National Natural Science Foundation of China (Nos. 12473047,12622307).

Mephisto is developed and operated by the South-Western Institute for Astronomy Research of Yunnan University (SWIFAR-YNU), funded by the ``Yunnan University Development Plan for World-Class University" and ``Yunnan University Development Plan for World-Class Astronomy Discipline".  
We thank the staff of IAO, Hanle, and CREST, Hosakote, that made HCT observations possible. The facilities at IAO and CREST are operated by the Indian Institute of Astrophysics, Bangalore. 
Based in part on observations collected at the European Organisation for Astronomical Research in the Southern Hemisphere, Chile, as part of ePESSTO+ (the advanced Public ESO Spectroscopic Survey for Transient Objects Survey -- PI: Inserra). ePESSTO+ observations were obtained under ESO program ID 112.25JQ.

This work has made use of data from the Asteroid Terrestrial-impact Last Alert System (ATLAS) project. The Asteroid Terrestrial-impact Last Alert System (ATLAS) project is primarily funded to search for near earth asteroids through NASA grants NN12AR55G, 80NSSC18K0284, and 80NSSC18K1575; byproducts of the NEO search include images and catalogs from the survey area. This work was partially funded by Kepler/K2 grant J1944/80NSSC19K0112 and HST GO-15889, and STFC grants ST/T000198/1 and ST/S006109/1. The ATLAS science products have been made possible through the contributions of the University of Hawaii Institute for Astronomy, the Queen's University Belfast, the Space Telescope Science Institute, the South African Astronomical Observatory, and The Millennium Institute of Astrophysics (MAS), Chile. We acknowledge the use of public data from the \textit{Swift} data archive.

This research has made use of the NASA/IPAC Extragalactic Database (NED), which is operated by the Jet Propulsion Laboratory, California Institute of Technology, under contract with NASA. We also acknowledge WISeREP - \url{https://www.wiserep.org}.
}
\facilities{Mephisto, Himalayan Chandra Telescope (HCT)}

\software{astropy \citep{astropy_2022ApJ...935..167A},  
          Matplotlib \citep{Matplotlib_2007}, 
          Numpy \citep{Numpy_2020},
          SciPy \citep{Scipy_2020},
          Photutils \citep{Photoutils_Bradley202514889440},
          GaiaXPy \citep{GAIAXPY_Daniela202517943031},
          AutoPhOT\citep{Brennan2022A&A...667A..62B},
          Source Extractor \citep{Bertin1996A&AS},
          SCAMP \citep{Bertin2010ascl.soft10063B},
          IRAF \citep{Tody1986SPIE},
          RedPipe \citep{Singh2021ascl.soft06024S},
          MOSFiT \citep{Guillochon2018ApJS..236....6G},
          dynesty \citep{dynesty_Speagle2020MNRAS.493.3132S},
          superbol \citep{Nicholl2018RNAAS},
          prospector \citep{Johnson2021ApJS..254...22J},
          HEASoft \citep{2014ascl.soft08004N}
          }

\bibliography{ref}{}
\bibliographystyle{aasjournal}

\appendix

\section{Properties of host galaxy}
\label{sec:app:host}
\begin{figure}[htbp]
\centering
\includegraphics[width=0.6\linewidth]{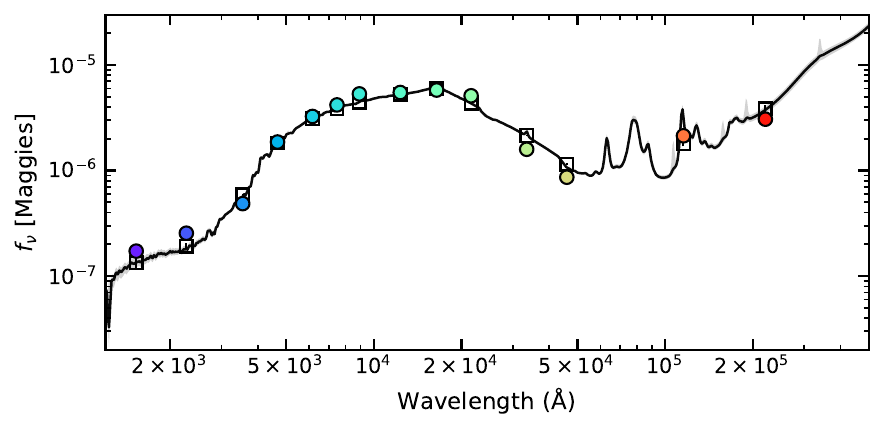}
\caption{SED of the host galaxy of SN~2025dys. Photometric points are marked in the coloured dots, while the best-fit model is present in the black solid lines. The grey region marks the 3-$\sigma$ confidence intervals.}
\label{fig:misc:hostsed}
\end{figure}

We utilized the \textsc{prospector} code \citep{Johnson2021ApJS..254...22J, Conroy2009ApJ...699..486C} to investigate the global properties of the host galaxy of SN~2025dys. The best-fit results are presented in Fig.~\ref{fig:misc:hostsed}. The modeling yields a total stellar mass of $M_{*}\approx10^{10}\,\mathrm{M_\odot}$, characterizing a moderately massive system that is slightly less massive than the Milky Way. The derived stellar metallicity is significantly sub-solar, at approximately $Z \approx 0.17\,\mathrm{Z_\odot}$. This metal-poor environment strongly suppresses mass loss via line-driven stellar winds, making binary interaction a far more plausible mechanism for the envelope stripping of the progenitor. Furthermore, the derived dust parameters, particularly a polycyclic aromatic hydrocarbon fraction of $Q_{\rm PAH}\approx1$, are entirely consistent with such a metal-deficient environment. Examining the star formation history, we observed a near-total cessation of star formation within the most recent $3\times10^{7}\,\mathrm{yr}$, exhibiting a rate of only $\sim 8\times10^{-3}\,\mathrm{M_\odot\,yr^{-1}}$. In stark contrast, during the epoch between $3\times10^{7}$ and $10^{8}\,\mathrm{yr}$ ago, the galaxy experienced a vigorous starburst, reaching a star formation rate of $\sim 2.55\,\mathrm{M_\odot\,yr^{-1}}$. This episodic history suggests that a galactic merger or interaction likely triggered a global starburst in the recent past, subsequently feeding the central supermassive black hole and igniting the Seyfert~2 active galactic nucleus \citep[see, e.g.,][]{Chen2022ApJS..258...29C}.

We note that SN~2025dys is located in the peripheral region of the host galaxy, at a projected distance of $\sim 11$~kpc from the galactic centre, which is far from any currently active star-forming regions. Crucially, the temporal gap between the historical starburst and the current epoch imposes a strict upper limit on the mass of the progenitor. An extremely massive star would have a lifespan significantly shorter than $3\times10^{7}\,\mathrm{yr}$; therefore, the progenitor likely a moderately massive star capable of surviving until the present day. A coherent physical picture thus emerges: the progenitor of SN~2025dys was born $30$ to $100$~Myr ago during the merger-induced starburst within a binary system. It subsequently evolved into a partially stripped configuration and either formed in situ within extended tidal structures triggered by the merger, or migrated outward to the galactic periphery due to complex merger dynamics prior to core collapse.

\section{Complementary tables and figures}
\subsection{Data reduction}\label{sec:app:ctf:red}

Promptly following the discovery of SN~2025dys, we initiated multi-band photometric monitoring using the \texttt{Mephisto} telescope in the $u_{\mathrm{M}}$, $v_{\mathrm{M}}$, $g_{\mathrm{M}}$, $r_{\mathrm{M}}$, $i_{\mathrm{M}}$, and $z_{\mathrm{M}}$ bands. The observations utilized the \texttt{BigCamera} instrument. All raw imaging data underwent standard reduction procedures within the \textsc{iraf} environment \citep{Tody1986SPIE}. These fundamental procedures included bias subtraction, dark current removal, flat-field correction, and field-of-view trimming. For epochs when the transient was faint, multiple exposures were obtained and subsequently combined via median stacking with sigma clipping to enhance the SNR. The pre-processed images were then analyzed using the dedicated \textsc{autophot} pipeline\footnote{\textsc{autophot} is a package for transient photometry using PSF fitting and template subtraction, see \url{https://github.com/Astro-Sean/autophot}} \citep{Brennan2022A&A...667A..62B}, following procedures similar to those described in \citet{SN2025aico_Zhao2026arXiv260710671Z}. Instrumental magnitudes and their associated uncertainties were extracted using point-spread-function (PSF) fitting techniques.

Final apparent magnitudes were calibrated against zero-points derived from unsaturated field stars cataloged in the Gaia database. The reference magnitudes and photometric uncertainties for these calibration stars were computed by performing synthetic photometry on the Gaia~XP spectra utilizing the \textsc{gaiaxpy}\footnote{\textsc{gaiaxpy} is a package to perform the calibration of Gaia XP spectra and synthetic photometry, see \url{https://gaia-dpci.github.io/GaiaXPy-website/}.} package \citep{GAIAXPY_Daniela202517943031}. Because the Gaia~XP spectra do not fully cover the wavelength ranges of the Mephisto $u_{\rm M}$ and $z_{\rm M}$ bands, an extrapolation technique was applied to both the blue and red extremities of the spectra, following the methodology detailed in \citet{Yang2024ApJ...969..126Y} and \citet{Chen2024ApJ...971L...2C}. Furthermore, we incorporated archival photometric data from public surveys. Light curves in the ATLAS cyan ($c$) and orange ($o$) bands were retrieved from the ATLAS Forced Photometry server\footnote{\url{https://fallingstar-data.com/forcedphot/}}. Additionally, space-based X-ray observations were downloaded from the \textit{Swift} data archive and processed using the \textsc{heasoft} suite \citep{2014ascl.soft08004N}.

We present a total of 13 epochs of optical spectroscopy for SN~2025dys, spanning from $+8.9$~days to $+74.0$~days after the explosion. The majority of the spectral sequence was obtained using the Himalayan Chandra Telescope (HCT) equipped with medium-resolution configurations (Gr\#7 and Gr\#8, $R \sim 1000$--$1200$), with high cadence density. Four additional spectra were retrieved from the European Southern Observatory (ESO) public archive (PI: Inserra, 112.25JQ), originally acquired with the NTT/EFOSC2 \citep[see,][]{Buzzoni1984Msngr..38....9B} instrument. Two supplementary spectra were sourced from the Transient Name Server (TNS\footnote{See \url{https://www.wis-tns.org/object/2025dys}}). A comprehensive summary of the spectroscopic observations is provided in Table~\ref{apptab:speclog}. The raw HCT/HFOSC data were reduced using the dedicated \textsc{redpipe}\footnote{\textsc{redpipe} is a package of scripts for data reduction for both imaging and spectroscopy, see \url{https://github.com/sPaMFouR/RedPipe}} software \citep{Singh2021ascl.soft06024S}. The basic reduction process of the HCT spectroscopic data included bias subtraction, overscan correction, flat-fielding, and image trimming. Following this, one-dimensional spectra were optimally extracted from the two-dimensional frames. Wavelength and flux calibrations were performed using arc lamp spectra and spectrophotometric standard stars, respectively, observed on the same night and with identical instrumental setups. Finally, the wavelength solutions were verified against prominent night-sky emission lines (e.g., [\ion{O}{i}]~$\lambda\lambda6300,\,6364$), and the absolute flux levels were scaled to match contemporaneous broad-band photometry. The data from the public archive, which were obtained from NTT/EFOSC2 are reduced by utilising the corresponding pipeline provided by ESO\footnote{See \url{https://www.eso.org/sci/software/pipe_aem_main.html}}.

\newpage
\subsection{Spectroscopic observation logs}
\label{sec:app:ctf:speclog}
\vspace{-1.0em}
\begin{table}[htbp]
\centering
\caption{Log of the spectroscopic observations of SN~2025dys.}
\label{apptab:speclog}
%\small
%\setlength{\tabcolsep}{5pt}
\begin{tabular}{cccccccc}
\hline
Date & MJD & Phase$^{a}$ & Instrumental setup & Grism/Grating & Spectral range & Exposure time & Resolution \\
(yyyymmdd) & & (d) & & & (\AA) & (s) & \\
\hline
20250317 & 60751.96 & $+$8.90  & NOT+ALFOSC  & Gr\#4          & 3390--9670  & \dots & $R \sim 360$ \\
20250318 & 60752.78 & $+$9.72  & HCT+HFOSC   & Gr\#7          & 3500--7500  & 2400  & $R \sim 1330$ \\
20250319 & 60753.67 & $+$10.61 & HCT+HFOSC   & Gr\#8          & 5800--9200  & 2400  & $R \sim 2190$ \\
20250328 & 60762.92 & $+$19.86 & CAHA+CAFOS  & G-200          & 3800--8700  & \dots & $R \sim 300$ \\
20250401 & 60766.79 & $+$23.73 & HCT+HFOSC   & Gr\#7$+$Gr\#8  & 3500--9200  & 2400  & $R \sim 1330$--$2190$ \\
20250404 & 60769.26 & $+$26.20 & NTT+EFOSC2  & Gr\#13         & 3400--10000 & 900   & $R \sim 350$ \\
20250414 & 60779.23 & $+$36.17 & NTT+EFOSC2  & Gr\#13         & 3400--10000 & 900   & $R \sim 350$ \\
20250415 & 60780.60 & $+$37.54 & HCT+HFOSC   & Gr\#7$+$Gr\#8  & 3500--9200  & 2480  & $R \sim 1330$--$2190$ \\
20250419 & 60784.65 & $+$41.59 & HCT+HFOSC   & Gr\#7$+$Gr\#8  & 3500--9200  & 1800  & $R \sim 1330$--$2190$ \\
20250429 & 60794.61 & $+$51.55 & HCT+HFOSC   & Gr\#7$+$Gr\#8  & 3500--9200  & 2700  & $R \sim 1330$--$2190$ \\
20250503 & 60798.18 & $+$55.12 & NTT+EFOSC2  & Gr\#13         & 3400--10000 & 1950  & $R \sim 350$ \\
20250511 & 60806.66 & $+$63.60 & HCT+HFOSC   & Gr\#7          & 3500--7500  & 2700  & $R \sim 1330$ \\
20250522 & 60817.09 & $+$74.03 & NTT+EFOSC2  & Gr\#13         & 3400--10000 & 1800  & $R \sim 350$ \\
\hline
\end{tabular}
\vspace{0.3em}
\begin{minipage}{\linewidth}
\footnotesize
$^{a}$Phases are relative to the estimated explosion epoch ($\mathrm{MJD} = 60743.06$) in the observer frame.
\end{minipage}
\end{table}
\vspace{-2.5em}

\subsection{Posteriors of light curve fitting}
\label{sec:app:ctf:mcmcpost}
\begin{table*}[htbp]
    \centering
    \caption{Multiband light curve fitting posterior results of SN~2025dys using \textsc{MOSFiT} with different power models.}
    \label{tab:mosfit_fitting_2025dys}
    \begin{tabular*}{\textwidth}{@{\extracolsep{\fill}} lcc c lcc}
        \hline
        \multicolumn{3}{c}{\textbf{RD}} & & \multicolumn{3}{c}{\textbf{Magnetar + RD}} \\
        \cline{1-3} \cline{5-7}
        Parameter & Unit & Posterior & & Parameter & Unit & Posterior \\
        \hline
        $f_{\mathrm{Ni}}$ & -- & $1.00$ & & $f_{\mathrm{Ni}}$ & -- & $0.07^{+0.09}_{-0.05}$ \\
        $M_{\mathrm{ej}}$ & $M_\odot$ & $0.62^{+0.03}_{-0.03}$ & & $M_{\mathrm{ej}}$ & $M_\odot$ & $0.20^{+0.04}_{-0.03}$ \\
        $v_{\mathrm{ej}}$ & $10^3\,\mathrm{km\,s^{-1}}$ & $8.13^{+0.38}_{-0.37}$ & & $v_{\mathrm{ej}}$ & $10^3\,\mathrm{km\,s^{-1}}$ & $5.37^{+0.13}_{-0.12}$ \\
        $t_{\mathrm{exp}}$ & $\mathrm{days}$ & $-2.83^{+0.25}_{-0.27}$ & & $t_{\mathrm{exp}}$ & $\mathrm{days}$ & $-1.55^{+0.10}_{-0.10}$ \\
        $\sigma$ & $\mathrm{mag}$ & $0.60^{+0.03}_{-0.03}$ & & $\sigma$ & $\mathrm{mag}$ & $0.25^{+0.01}_{-0.01}$ \\
         &  &  & & $B$ & $10^{14}\,\mathrm{G}$ & $4.68^{+0.57}_{-0.51}$ \\
         &  &  & & $P_{\mathrm{spin}}$ & $\mathrm{ms}$ & $13.18^{+0.31}_{-0.30}$ \\
        \hline
        \multicolumn{3}{c}{\textbf{CSI + RD ($s=0$)}} & & \multicolumn{3}{c}{\textbf{CSI + RD ($s=2$)}} \\
        \cline{1-3} \cline{5-7}
        Parameter & Unit & Posterior & & Parameter & Unit & Posterior \\
        \hline
        $f_{\mathrm{Ni}}$ & -- & $0.017^{+0.003}_{-0.002}$ & & $f_{\mathrm{Ni}}$ & -- & $0.015^{+0.003}_{-0.002}$ \\
        $E_{\mathrm{k}}$ & $10^{51}\,\mathrm{erg}$ & $0.37^{+0.06}_{-0.06}$ & & $E_{\mathrm{k}}$ & $10^{51}\,\mathrm{erg}$ & $0.46^{+0.04}_{-0.06}$ \\
        $M_{\mathrm{CSM}}$ & $M_\odot$ & $0.76^{+0.13}_{-0.10}$ & & $M_{\mathrm{CSM}}$ & $M_\odot$ & $0.54^{+0.05}_{-0.02}$ \\
        $M_{\mathrm{ej}}$ & $M_\odot$ & $2.45^{+0.30}_{-0.37}$ & & $M_{\mathrm{ej}}$ & $M_\odot$ & $2.69^{+0.33}_{-0.35}$ \\
        $R_0$ & $\mathrm{AU}$ & $83^{+17}_{-12}$ & & $R_0$ & $\mathrm{AU}$ & $6.2^{+9.7}_{-4.8}$ \\
        $\rho$ & $10^{-12}\,\mathrm{g\,cm^{-3}}$ & $1.82^{+0.52}_{-0.41}$ & & $\rho$ & $10^{-11}\,\mathrm{g\,cm^{-3}}$ & $1.17^{+23}_{-0.97}$ \\
        $t_{\mathrm{exp}}$ & $\mathrm{days}$ & $-2.07^{+0.18}_{-0.17}$ & & $t_{\mathrm{exp}}$ & $\mathrm{days}$ & $-0.75^{+0.07}_{-0.07}$ \\
        $\sigma$ & $\mathrm{mag}$ & $0.22^{+0.01}_{-0.01}$ & & $\sigma$ & $\mathrm{mag}$ & $0.29^{+0.01}_{-0.01}$ \\
        \hline
    \end{tabular*}
    
    \begin{minipage}{0.98\textwidth}
        \footnotesize
        \textbf{Notes.} Posterior values represent the median with $16\mathrm{th}$ and $84\mathrm{th}$ percentiles converted to linear scale. All logarithmic parameters from the original fits have been transformed via $10^x$. For the RD model, $f_{\mathrm{Ni}}$ is pegged at the upper prior boundary ($1.0$). $t_{\mathrm{exp}}$ is the explosion epoch relative to the first observation. $\sigma$ corresponds to the additional variance added in quadrature to the photometric uncertainties. $R_0$ and $\rho$ represent the inner radius and reference density of the CSM, respectively.
    \end{minipage}
\end{table*}

\end{document}